\def\year{2022}\relax
\documentclass[letterpaper]{article} % DO NOT CHANGE THIS
\usepackage{aaai22}  % DO NOT CHANGE THIS
\usepackage{times}  % DO NOT CHANGE THIS
\usepackage{helvet}  % DO NOT CHANGE THIS
\usepackage{courier}  % DO NOT CHANGE THIS
\usepackage[hyphens]{url}  % DO NOT CHANGE THIS
\usepackage{graphicx} % DO NOT CHANGE THIS
\usepackage{natbib}  % DO NOT CHANGE THIS AND DO NOT ADD ANY OPTIONS TO IT
\usepackage{caption} % DO NOT CHANGE THIS AND DO NOT ADD ANY OPTIONS TO IT
\DeclareCaptionStyle{ruled}{labelfont=normalfont,labelsep=colon,strut=off} % DO NOT CHANGE THIS
\usepackage{algorithm}
\usepackage{algorithmic}

\usepackage{newfloat}
\usepackage{listings}
\floatstyle{ruled}
\newfloat{listing}{tb}{lst}{}
\floatname{listing}{Listing}
\title{COVID-19 Vaccine Misinformation Campaigns and Social Media Narratives}
\author {
    Karishma Sharma,
    Yizhou Zhang,
    Yan Liu
}
\affiliations {
    University of Southern California \\
    krsharma@usc.edu, zhangyiz@usc.edu, yanliu.cs@usc.edu
}

\usepackage{enumitem}
\usepackage{subcaption}
\usepackage{booktabs}

\begin{document}

\maketitle

\begin{abstract}
COVID-19 vaccine hesitancy has increased concerns about vaccine uptake required to overcome the pandemic and protect public health. A critical factor associated with anti-vaccine attitudes is the information shared on social media. In this work, we investigate  misinformation communities and narratives that can contribute to COVID-19 vaccine hesitancy.

During the pandemic, anti-science and political misinformation/conspiracies have been rampant on social media. Therefore, we investigate misinformation and conspiracy groups and their characteristic
behaviours in Twitter data collected on COVID-19 vaccines. We identify if any suspicious coordinated efforts are present in promoting vaccine misinformation, and find two suspicious groups - one promoting a ‘Great Reset’ conspiracy which suggests that the pandemic is orchestrated by world leaders to take control of the economy, with vaccine related misinformation and strong anti-vaccine and anti-social messages such as no lock-downs; and another promoting the Bioweapon theory. Misinformation promoted is largely from the anti-vaccine and far-right communities in the 3-core of the retweet graph, with its tweets proportion of conspiracy and questionable sources to reliable sources being much higher. In comparison with the mainstream and health news, the right-leaning community is more influenced by the anti-vaccine and far-right
communities, which is also reflected in the disparate vaccination rates in left and right U.S. states. The misinformation communities are also more vocal, either in vaccine or other discussions, relative to remaining communities, besides other behavioral differences. 

Furthermore, we investigate the COVID-19 vaccine narratives spread on social media. Besides misinformation narratives about vaccine safety, effectiveness and conspiracies, we find that \emph{rarer} vaccine side-effects, reported less frequently in CDC VAERS reports, were more frequently discussed on social media, and in misinformation narratives, which also use other known tactics of science narratives distortion. 
\end{abstract}

\section{Introduction}
\label{intro}

The COVID-19 pandemic has amplified the concerns surrounding social media communications, which have on one hand facilitated pro-social messaging like ``\#wearamask", ``\#staysafe" \cite{SocialMediaCovidRole}, but also has been a breeding ground for health and political misinformation and conspiracies \cite{sharma2020covid}. While global efforts have been made to rapidly vaccinate people against COVID-19 and prevent the risk of severe illness, there is significant unwillingness to vaccinate in part of the population \cite{massonvaccinate}. % reported a quarter of American adults said they would refuse the vaccine, and 5\% were undecided.

% Vaccine hesitancy and misinformation on social media and e-commerce platforms has gained much attention in the past few years \cite{cossard2020falling,juneja2021auditing}. \cite{cossard2020falling} studied the Italian vaccine debate finding echo chambers of anti-vaccine and pro-vaccine groups in 2016 on Twitter, with interaction between the communities being asymmetrical, as vaccine advocates ignore the skeptics. Similarly, \cite{miyazaki2021strategy} found anti-vaccine accounts replying most to neutral accounts using toxic and emotional content. Studies have found the growing debate about the merits of vaccination on social media to be accompanied by reduced vaccinations, reduced intent to vaccinate and reappearance of diseases like measles \cite{smith2011parental,cossard2020falling,ortiz2019systematic}.

% During COVID-19, misinformation, conspiracies and coordinated misinformation campaigns, have been highly prevalent throughout the pandemic \cite{sharma2021identifying,memon2020characterizing}. \cite{sharma2021identifying} found coordinated accounts promoting political conspiracies and anti-mask narratives, \cite{memon2020characterizing} characterized health misinformation about COVID-19, among other similar studies. Given the widespread misinformation, government agencies such as the CDC, and fact-checking websites, and communication studies \cite{lewandowsky2021covid} have curated lists of COVID-19 vaccine myths on their websites to inoculate the public against vaccine misinformation.

Earlier COVID-19 misinformation studies found narratives related to cures, prevention, mortality rate, hoax, scientific facts, and partisan conspiracies \cite{memon2020characterizing,jiang2020political}, both diminishing the seriousness of the virus, and opposing public health measures with anti-mask, anti-lockdown campaigns \cite{sharma2021identifying}. With the availability of COVID-19 vaccines, it has similarly become a target for misinformation. An earlier study found that participants shown material supporting anti-vaccine conspiracies exhibited lower vaccination intent than those in anti-conspiracy conditions or controls \cite{jolley2014effects}.

Since differential exposure to vaccine misinformation and scientific health information can be detrimental \cite{jolley2014effects}, we focus on characterization of misinformation and information communities in the COVID-19 vaccine discussion. In this work, we characterize suspicious hidden efforts in coordinated promotion of misinformation, anti-vaccine conspiracy communities, and information and health news communities along with their narratives. The key findings are that (i) anti-vaccine attitudes in the COVID-19 vaccine discussion are correlated with partisan attitudes, with right-leaning communities closer to retweeting content from anti-vaccine misinformation communities (ii) misinformation and coordinated conspiracies promote distrust in public health authorities, and COVID-19 vaccines safety and effectiveness. (iii) distortion of facts with false, misleading and conspiracy narratives encompass scientific facts, safety and effectiveness, and political conspiracies, including more frequent discussion of rarer side-effects on social media.

Recent ongoing studies on social media discourse about COVID-19 vaccines, and past studies on vaccines in general, and our work share some overlapping and new insights. \citeauthor{cossard2020falling} \citeyear{cossard2020falling} findings of anti-vaccine and pro-vaccine echo-chambers in Italian social media vaccine debate reaffirm our findings of anti-vaccine Italian communities in the larger COVID-19 vaccine discussion, along with presence of similar echo-chamber structures based on partisan ideology and conspiracy groups (far-right conspiracy group, and a larger anti-vaccine community of English tweets, besides French, Italian and Spanish-En COVID-19 anti-vaccine communities).

With inferred political ideology, geographical analysis, misinformation sources proportion shared by each community, and account characteristics, we provide detailed insights about the misinformation and informational communities and their structure in the discourse. In recent concurrent work, \citeauthor{miyazaki2021strategy} \citeyear{miyazaki2021strategy} found that COVID-19 anti-vaccine accounts interact through toxic replies to pro-vaccine accounts, while the latter ignore the former. In another concurrent work, \citeauthor{pierri2021impact} \citeyear{pierri2021impact} also confirms the negative correlation between misinformation rate and vaccination rate by partisan nature of US states as in our findings.
In addition to the previous findings, we found suspicious coordinated efforts behind two conspiracies - suggesting a `Bioweapon' or Dehumanization theory, and a `Great Reset' political conspiracy with greater automated and colluding account behaviours. We also observed that distortion of facts with misleading rather than outright false narratives, through exaggeration of rarer vaccine side-effects, questioning of scientific facts, and coordinated promotion of conspiracies are present, which can be more challenging for misinformation detection and mitigation efforts. Our insights can bring attention to the nuances of distorted facts to further research in detection and contextualization of social media biases and misleading narratives from anti-vaccine groups, and mitigation efforts to prevent coordinated conspiracies, and to reduce the partisan and anti-vaccine misinformation.

We examine the following research questions:

% RQ1. Are there hidden coordinated efforts promoting misinformation/conspiracies about COVID-19 vaccines? What other misinformation communities are present, and which communities are most influenced by them?

% most to neutral accounts using toxic and emotional content. % , and the effects are different across different demographics. Moreover, \cite{jamison2020not} also found that vaccine opponents share greater proportions of unreliable information. % \cite{pierri2021impact} reported higher hesitancy and misinformation in Republican counties, but larger effects or change in hesitancy with misinformation rate in Democratic counties.

% In this work, we uncover and characterize misinformation campaigns and communities in the COVID-19 vaccine discussion on Twitter and examine their anti-vaccine propaganda towards increasing vaccine hesitancy and reducing vaccination uptakes in the US. 

% Does social media contribute to COVID-19 vaccine hesitancy through vaccine misinformation promoted by coordinated misinformation campaigns and anti-vaccine misinformation communities?
\begin{itemize}[wide=0pt]
    \item \textbf{RQ1. Are there hidden coordinated efforts promoting misinformation/conspiracies about COVID-19 vaccines? What misinformation communities are present, and which communities are most influenced by them?} {We apply AMDN-HAGE \cite{sharma2021identifying} to identify hidden coordinated efforts in misinformation promotion from observed account activities, and we apply Louvain method \cite{blondel2008fast} for community detection on the retweet graph. We characterize  communities by top retweeted accounts, tweet features, political leaning, geographical demographic, and misinformation proportion.}
    
    We find suspicious coordinated groups, one promoting a ‘Great Reset’  conspiracy  which  suggests  that  the  pandemic  is  orchestrated by world leaders to take control of the economy,with anti-vaccine/lockdown misinformation, and another promoting the Bioweapon theory. In the 3-core of the retweet graph, we find a large anti-vaccine misinformation and conspiracy community (16\%) that spans US (48.9\%) and UK (27.5\%) accounts, and other smaller anti-vaccine communities of different languages (Spanish, French, Italian). Misinformation promoted on the network is largely from the anti-vaccine and far-right communities, with its tweets' proportion of conspiracy and questionable to reliable sources being much higher at about $60\%$. The right-leaning community (that retweets top Republican accounts) are closer to the anti-vaccine and far-right conspiracy communities, compared to mainstream and health news, correlated with lower vaccination rates in Republican states.

\item \textbf{RQ2. How do the behaviours of anti-vaccine misinformation/conspiracy communities differ from informational communities, in general and specific to the vaccine discussion?} {We investigate distribution of account features in the communities to characterize their behaviours.}

We find that the anti-vaccine misinformation community is the most vocal in the vaccine discussion even though it has fewer tweets in totality (considering tweets outside of the vaccine discussions), has younger account ages than other communities, and it has a sizable audience. % , and more decentralized than hub structure. 
% In contrast, the far-right conspiracy group is also more vocal but in other discussions compared to the vaccine discussions
In contrast, the far-right conspiracy group is more vocal but in other discussions compared to the vaccine discussions, and is more interconnected with accounts having more followers-followings. In the vaccine discussion, mainstream news and left-leaning communities are more vocal than far-right and right-leaning ones, but less  than the anti-vaccine community.

\item \textbf{RQ3. What are the misinformation narratives present and spread through social media ?} {We consider five types of science narratives distortions and examine how they manifest on social media through correlation of reported vaccine side-effects with frequency in the tweets, topic modeling, and news source characterization for misinformation tweets.}

We observe misinformation narratives distort facts through false, misleading, and conspiracy claims. We find that rarer side-effects are discussed more frequently in tweets and misinformation narratives. Other tactics such as setting impossible expectations about vaccines effectiveness and scientific facts, logical fallacies such as vaccines are rushed or experimental, pseudoscience, political propaganda, and conspiracies such as the pandemic is planned or a `Great Reset' or vaccines are a Bioweapon for dehumanization, were present.

% We find that rarer  Other tactics to distort facts through misleading, false information, and conspiracies are present, besides cherry-picking, such as setting impossible expectations of vaccines, logical fallacies such as rushed or experimental process, factual inaccuracies, pseudoscience, and political propaganda. 

% We find that rarer vaccine side-effects are discussed more frequently in all tweets, and it is further amplified in misinformation tweets, explained by novelty and cherry-picking of events. Also, misinformation tweets had used tactics besides cherry-picking, such as setting impossible expectations of vaccines, conspiracies, misleading about vaccines being experimental or rushed, and other factual inaccuracies, pseudoscience and political propaganda, all of which can contribute to increased vaccine hesitancy.
% and the pandemic provides an opportunity for reset. The group promotes baseless COVID-19 anti-vaccine misinformation and conspiracies, with anti-social messages (\#notocoronavirusvaccines, \#livingnotlockdown, \#masksdontwork).
\end{itemize}

% \begin{enumerate}
%     \item \textb
%     \item \textbf{RQ1}: Are retweet communities of {anti-vaccine and misinformation or conspiracy groups} on Twitter distinct from general accounts? How segregated are the information propagation dynamics of retweet communities? Is there presence of {coordinated misinformation campaigns}; what are its narratives and target communities?
%     \item \textbf{RQ2}: Do the network dynamics i.e., structure and engagements of Twitter communities {evolve} over time?
%     \item \textbf{RQ3}: What strategies and manipulation tactics do the anti-vaccine and misinformation or conspiracy groups use to promote vaccine hesitancy? How much is the correlation between vaccine side effects officially reported in CDC VAERS vs discussed on social media, and can it potentially lead to increased hesitancy?
%     % \item \textbf{RQ4}: What are the characteristic behaviours of these accounts at the start of the COVID-19 pandemic? % Can we build a classifier to identify accounts who might be susceptible to developing an anti-vaccine stance?
% \end{enumerate}

\section{Data Collection}

\subsubsection{COVID-19 Vaccine Twitter Data.} 
\begin{figure}[t]
    \centering
    \includegraphics[width=\columnwidth,height=4cm]{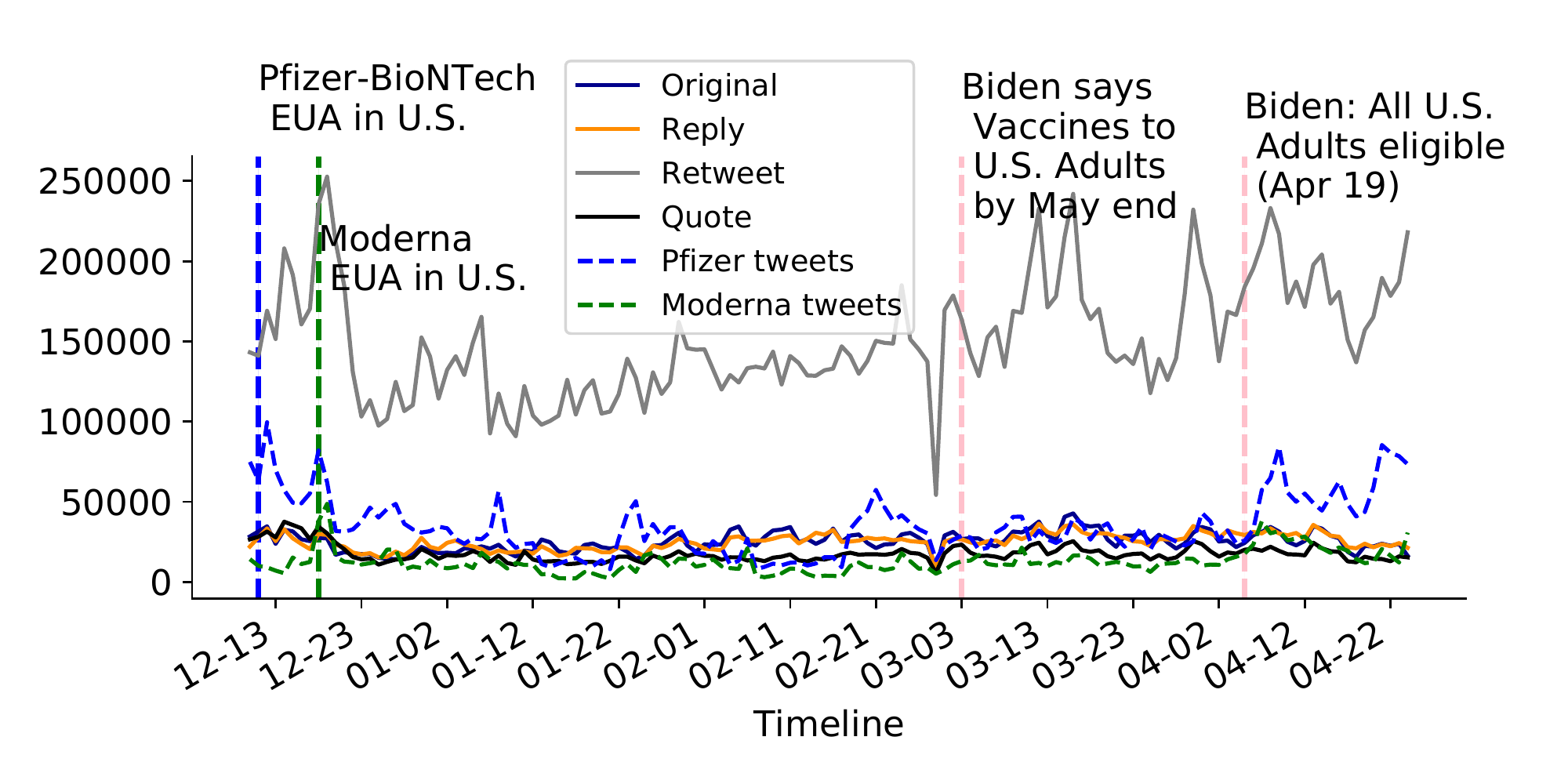}
    \caption{Tweet volume timeline from the Emergency Use Authorization (EUA) in U.S. till U.S. Adults vaccinations.}
    \label{fig:vol_timeline}
\end{figure}

We use the streaming Twitter API which returns a $\sim$1\% sample of all tweets filtered by tracked keywords (vaccine, Pfizer, BioNTech, Moderna, Janssen, AstraZeneca, Sinopharm) in real-time, to collect Twitter data related to COVID-19 vaccines. The dataset collection includes tweets from Dec 9, 2020 - April 24, 2021%\footnote{For dates between 2020-01-12 - 2021-01-22, and 2021-02-04 - 2021-02-16, due to technical interruptions, we recover some tweets from CoVaxxy \cite{deverna2021covaxxy}, which does not contain tweets in December after EUA of vaccines.}
, i.e., just before Pfizer-BioNTech and Moderna were approved by the FDA for Emergency Use Authorization (EUA). The dataset contains 29,743,178 tweets from 7,417,592 accounts. Fig~\ref{fig:vol_timeline} shows the timeline of tweets.\footnote{https://www.ajmc.com/view/a-timeline-of-covid-19-vaccine}

\subsubsection{CDC-VAERS U.S. Post-Vaccine Side-Effects Records.\footnote{https://vaers.hhs.gov/data.html}} Post vaccination side-effects are reported to the FDA/CDC Vaccine Adverse Event Reporting System
(VAERS). We download the official records from (accessed June 6, 2021).
Healthcare providers are required to report, while individuals are advised to report post vaccine effects, even if a causal link to the vaccine has not yet been established for monitoring. We use the data to study correlation between discussion of side-effects in VAERS and on social media. % , in order for the FDA/CDC to monitor vaccine safety and conduct further investigations. The reports do not provide exact statistics of vaccine side-effects, but a public account of nation-wide post vaccine effects. We use the data as an estimate to study how post vaccine experiences are discussed on social media in comparison to ones reported to VAERS. % incomplete, inaccurate, coincidental, or unverifiable

\subsubsection{Unreliable/Conspiracy News-Source Credibility Lists.} We use misinformation as an umbrella term to refer to unreliable (false, misleading, and conspiracy) claims. Low-quality news-sources to analyze misinformation shared on social media are used in numerous prior works \cite{Bozarth_Saraf_Budak_2020}. % and it is found that temporal trends and differences in topics between misinformation and mainstream news were robust to the choice of news source . 
We utilize low-quality sources reported by three fact-checking resources, as consistently promoting COVID-19 or general misinformation: Media Bias/Fact Check (questionable and pseudoscience/conspiracy lists with low/very low factual rating), NewsGuard (accessed September 22, 2020),
and  Zimdars \cite{zimdars2016false} tagged as unreliable or related labels. For reliable mainstream news sources, Wikipedia:Reliable sources/Perennial sources tagged as reliable are included. In total, we obtain 124 reliable and 1380 unreliable/conspiracy sources.

\section{Coordinated Misinformation Campaigns} % about COVID-19 Vaccines}
In recent years, social media has witnessed misinformation or influence campaigns from network of both human and bot accounts colluding maliciously to promote specific agendas and misinformation \cite{sharma2021identifying,zhang2021vigdet}. Different from social bots, or accounts sharing similar contents or topical interests, uncovering suspicious coordinated groups is a hard task since the mechanism of coordination employed is unknown, and inconsistent over time. A recent unsupervised method AMDN-HAGE \cite{sharma2021identifying} is a significant advance over earlier approaches in estimating coordination efforts from observed activity traces on social media and can also be integrated with prior domain knowledge \cite{zhang2021vigdet}. \citeauthor{sharma2021identifying} \citeyear{sharma2021identifying} evaluated the method discussed below on Russian coordinated campaigns in US 2016 Election data with ground-truth labels from the US Congress investigations, achieving 0.93 AUC and 0.73 F1 having a large margin over the existing unsupervised methods.

\begin{figure*}[t]
    \centering
    \includegraphics[scale=0.4]{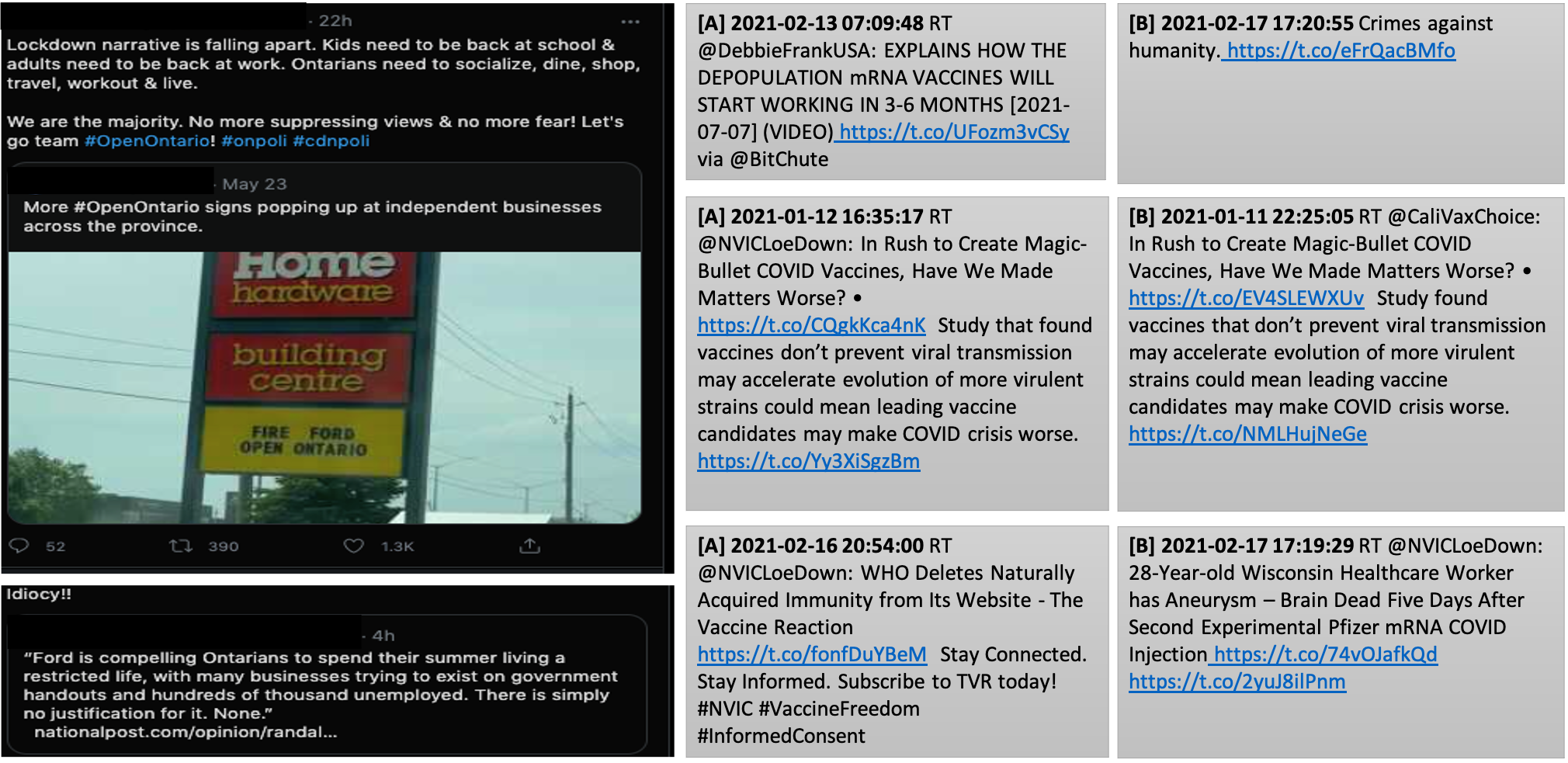}
    \caption{\textbf{Tweets from a pair of accounts (A, B) in the detected coordinated group.} Left: Tweets from the Twitter profile of accounts A and B suggesting anti-lockdown and anti-government narratives. Right: Three example tweets from the collected dataset, of the same pair of accounts (A, B) suspected of amplifying misinformation by coordinatedly sharing similar agendas.}
    \label{fig:coord_tweets}
\end{figure*}

\begin{figure*}[t]
    \centering
    \includegraphics[scale=0.4]{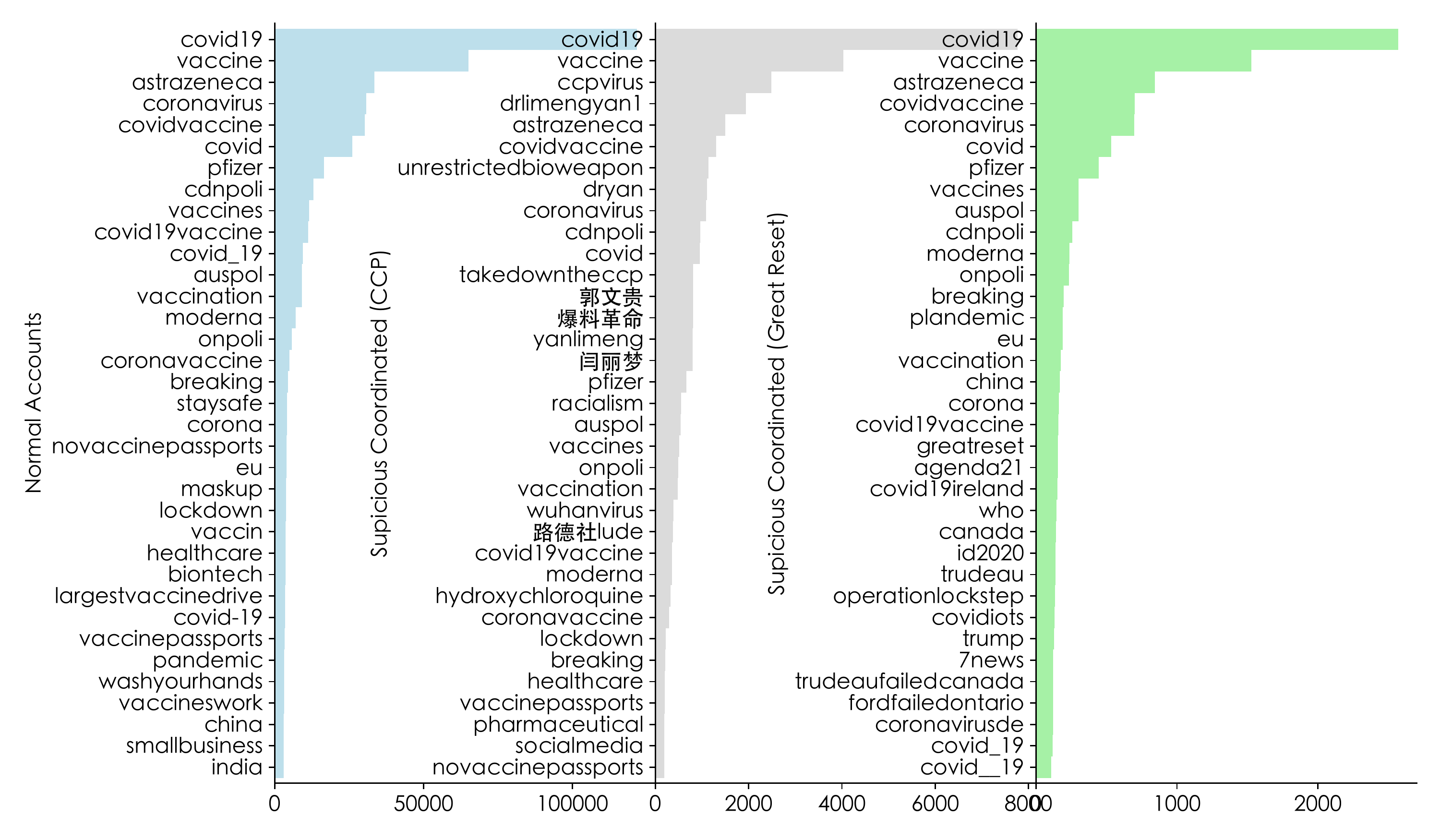}
    \caption{Top-35 hashtags of normal and identified suspicious coordinated accounts. Unique in each group in bold.}
    \label{fig:coordinated_hashtags}
\end{figure*}

\begin{figure}[t]
    \centering
    \includegraphics[width=\columnwidth,height=5cm]{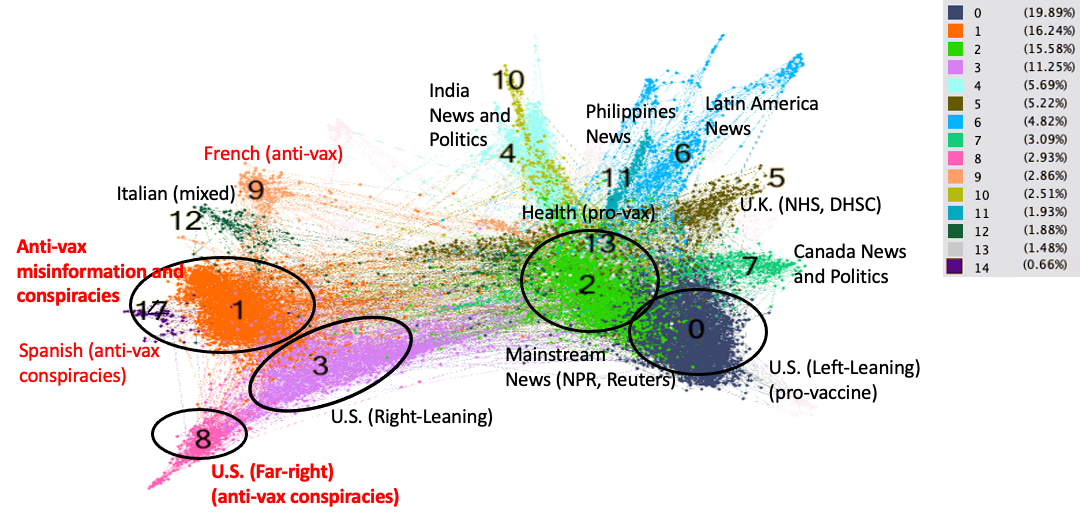}
    \caption{Characterization of communities in 3-core of Retweet (RT) graph to find misinformation communities.}
    \label{fig:RT_coms}
\end{figure}

\subsection{Methodology} We apply unsupervised AMDN-HAGE \cite{sharma2021identifying} to identify suspicious coordinated accounts in the dataset. 
% Misinformation campaigns involving coordinated efforts from groups of malicious accounts towards manipulating public opinion have become increasingly prevalent. % Here, we apply a variant of the method which can additionally encode domain information to regularize the learning from the observed data, i.e., higher co-occurrence frequencies and overlapping periods of activity between account pairs which might suggest higher likelihood of coordination between the accounts.
AMDN-HAGE models hidden influence between accounts activities from observed activity traces (account ids and timestamps). Specifically, it models the conditional density of future activities given past activities, thereby estimating a \textbf{mutual triggering effect (influence) between accounts}. The model parameters to estimate the conditional density (and influence) are trained using maximum likelihood (MLE) from observed social media data, as an unsupervised generative model. AMDN-HAGE assumes that coordinating accounts have a (1) hidden influence between their activities (centrally controlled, externally colluding, or jointly motivated) (2) are collectively anomalous from other accounts that are not colluding to promote. 

Other accounts who are operating individually (not in coordination) would have less organized or more independent and randomized activity patterns. Leveraging this, AMDN-HAGE estimates this mutual account influence and group behavior to uncover coordinated groups directly from observed activities. It does not observe or utilize any content or tweet features (other than the time-stamped sequence of activities from accounts) and therefore captures coordination behaviors rather than topical interests of accounts. Methods like Louvain community detection \cite{blondel2008fast} have distinct differences that make them less suitable for detecting hidden coordination of collaborating accounts (1) Community detection would capture direct interactions e.g. retweet graph communities. However organized campaign accounts might operate through a hidden indirect influence rather than direct retweets of each other. (2) Topical interest based associations from a retweet graph or other pre-defined graphs in community detection result in many false positives \cite{sharma2021identifying,pacheco2020uncovering}.

\subsubsection{Set-Up.} We construct observed sequences of accounts' activities from the diffusion cascade of a tweet, i.e., sequence of retweet, reply, quotes of the tweet (direct engagements) and all subsequent engagements to those, as a time-ordered sequence of posts represented as $C=((u_i, t_i))_{i=1}^L$ corresponding to account $u_i$, posting time $t_i$, and number of posts L i.e., sequence length. The extracted sequences contain 316k activity cascades of 205k accounts, after filtering accounts less than 20 times in the collected tweets, and cascade sequences shorter than length 5. % (Dec - Feb 24). 
We applied the method\footnote{% The model was trained on 4 Nvidia-2080Ti GPUs with account embedding dimensions set to 64 implemented in PyTorch with Adam optimizer (1e-3 learning rate, 1e-5 regularization) as in \cite{sharma2021identifying}
The AMDN-HAGE implementation is available at  https://github.com/USC-Melady/AMDN-HAGE-KDD21.} to the observed sequences to identify coordinated account groups, which resulted in 3 clusters (the method found 2 distinct small group of accounts that are suspicious of coordination, and the large group is the rest of the accounts referenced here as `Normal' i.e., non-coordinated as estimated by the method.) where the silhouette score has highest increase with the max at 10 clusters which are at finer granularity. 

\subsection{Analysis} We examine the two identified account groups ($\sim$8k and 3k accounts) and the remaining `Normal' accounts in terms of tweets features and account behaviors. The tweets from the identified coordinated group contained 5\% more misinformation (unreliable/conspiracy URLs in tweets) than over all tweets. Although there were false positives due to the large scale of accounts which makes clustering and learning harder, the groups identified were notably suspicious in terms of the content promoted in their tweets, even though the model has never seen the tweet contents.  

\subsubsection{Tweet Features and Conspiracies.} In Fig.~\ref{fig:coordinated_hashtags}, we compare the lowercase top-35 hashtags in the tweets of each group (in bold are the non-overlapping hashtags).
The coordinated conspiracy group narratives focus on the pandemic being a hoax (`plandemic'), with one group focusing on the `Great Reset' conspiracy, including `Agenda 21', `Operation LockStep` which are spin-offs of real-world projects, to falsely suggest malicious intents of world leaders in planning the pandemic towards global economic control.\footnote{https://www.bbc.com/news/55017002} The conspiracy started trending globally after a video of Canadian Prime Minister Justin Trudeau at a UN meeting talking about economic recovery or reset went viral. 
Tweets from the second coordinated group promote the Bioweapon theory that the virus is a Chinese (CCP) originated Bioweapon. Both coordinated group tweets contain anti-vaccine misinformation. The top hashtags in normal accounts support health interventions (\#maskup, \#healthcare, \#staysafe). Example vaccine misinformation tweet from the coordinated group,

\begin{quote}
``@CNN Remember the Covid vaccine is substantially more dangerous than the virus. Issues range from severe allergic reactions to blindness, stroke and even sudden death! You have been warned! \#Plandemic \#Agenda21 \#ID2020 \#Operationlockstep \#covidvaccine \#Coronavirus \#Covid19 \#Greatreset."
\end{quote}

% \small\emph{``@Newsweek World War V5.0: Covid virus is a bio weapon made by CCP.  Vaccine is a part of plan to wipe out all white people. \#DrLiMengYan1 \#CCPVirus \#Covid19 \#TakeDownTheCCP."}

% We examine tweets of the identified coordinated group with Twitter suspensions 4.26\% (vs. 3.45\% in normal group), and recently created accounts in 2021 19\% (vs. 15\% in normal group), although misinformation/suspensions are not exclusive to coordinated groups). % although misinformation is not exclusive to coordinated groups). % compared to reliable urls (100-52 is reliable).
% . 
% Examples of account profiles (Fig~\ref{fig:coord_account_profiles}) from the detected coordinated group shows the `Great Reset' conspiracy and nature of tweets, which is strongly anti-vaccine, anti-lockdown
\subsubsection{Account Activities.} 
We inspected activities of a sample of the detected coordinated accounts. We randomly sampled account pairs that had retweeted at least one common tweet in the observed collected dataset. For a pair of accounts, we checked their Twitter profile and their tweets in the collected dataset. Fig.~\ref{fig:coord_tweets} shows an example account pair (A, B) from the coordinated group, still active on Twitter as of June, 2021. The account names are anonymized here. The tweets of both accounts promoted the same agendas in coordination over similar time periods. In one instance, both retweet different sources that independently posted the same content, seemingly part of a coordinated network, as shown in the example @NVICLoeDown and @CaliVaxChoice posted exactly the same content and they re-shared each respectively.

\begin{figure}
    \centering
    \includegraphics[scale=0.4]{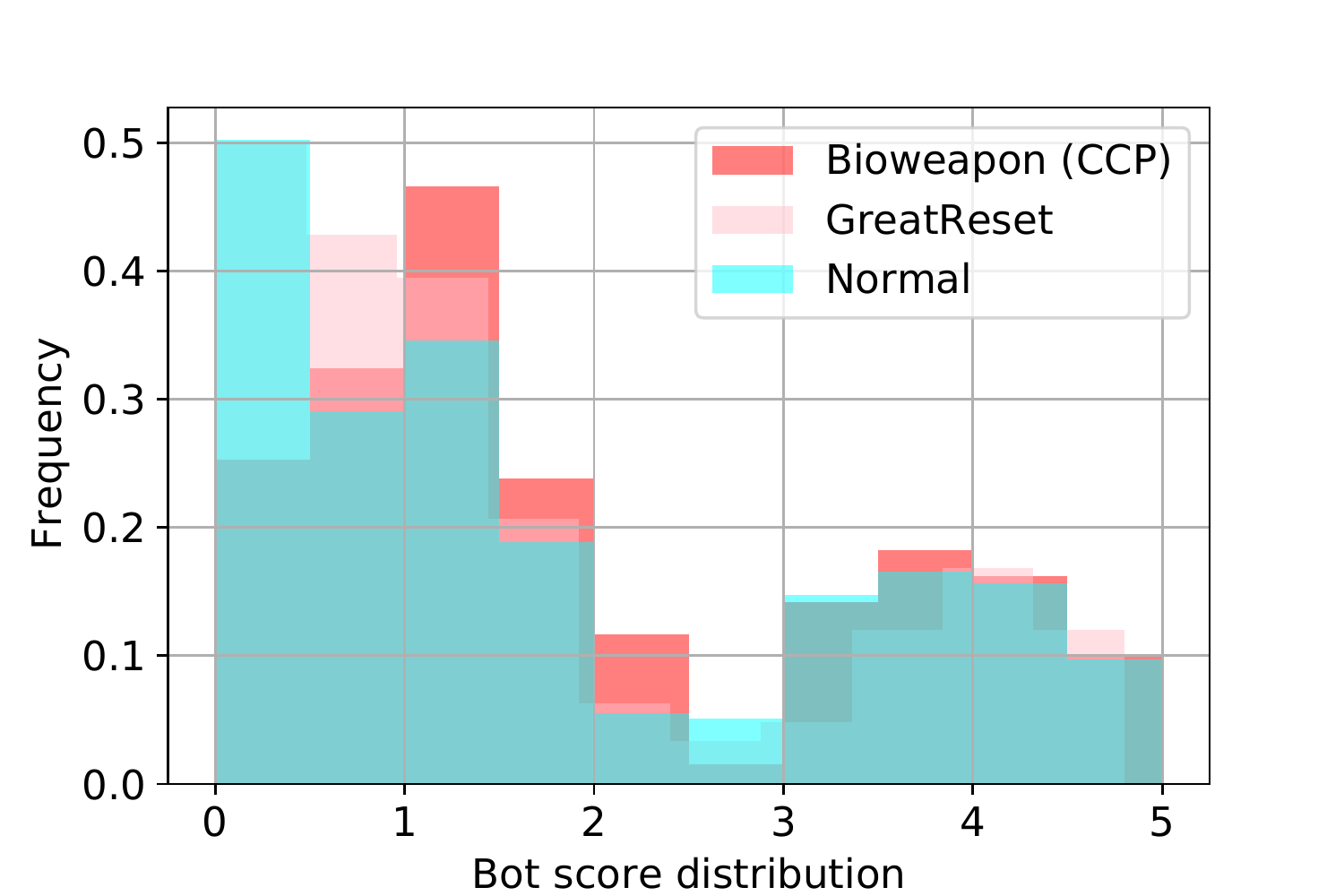}
    \caption{Bot score distribution. Mann-Whitney U-Test for suspicious coordinated Bioweapon (CCP) vs. Normal accounts sample (z-score -2.56, p-val 0.00523 $<$ 0.05) and suspicious coordinated (Great Reset) vs Normal accounts sample (z-score -1.35, p-val 0.0869 $<$ 0.1).}
    \label{fig:bot_scores}
\end{figure}

\begin{figure}
    \centering
    \includegraphics[scale=0.4]{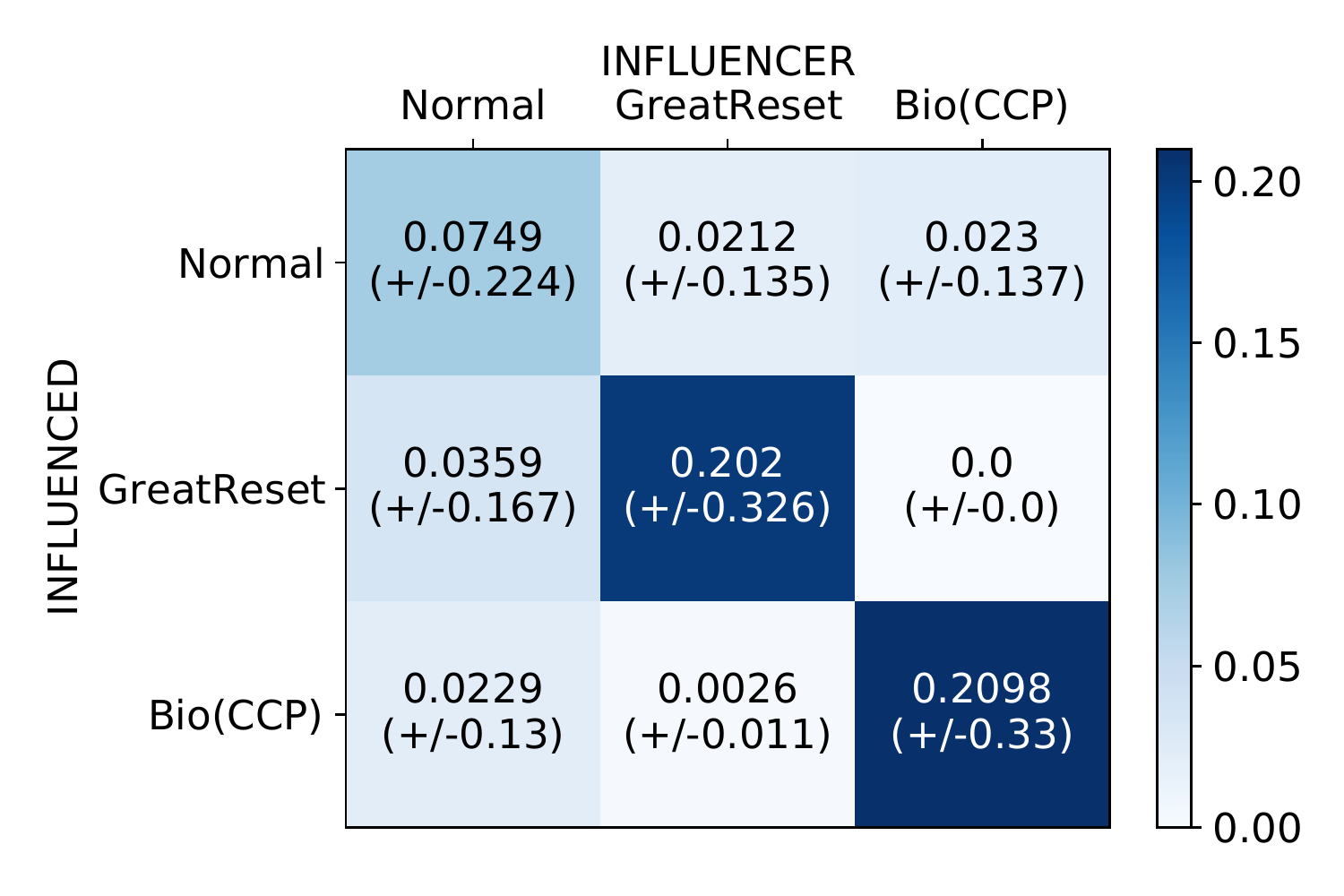}
    \caption{Mutual triggering effect (influence). Between activities of accounts, estimated from data by AMDN-HAGE shown as Avg. estimated triggering effect from Influencer accounts (whose activities trigger future activities in time). Normal accounts have weaker influence patterns (more random activities) compared to coordinating accounts.}
    %Average estimated triggering effect from Influencer accounts for detected suspicious coordinated and normal accounts sample.}
    \label{fig:infl_coordinated}
\end{figure}

\subsubsection{Quantitative Analysis.} \textbf{Fig~\ref{fig:bot_scores}} examines the \textbf{bot i.e., automated account score distribution} of suspicious coordinated group accounts. Conspiracies promoted from coordinating account groups (colluding in a hidden, unknown manner) tend to employ both bot (automated) and human actors to push agendas \cite{luceri2020detecting,badawy2019characterizing}. We evaluate the bot scores using Botometer v4 API \cite{sayyadiharikandeh2020detection} on 500 randomly sampled accounts each from normal and coordinated groups to compare the distribution. We assume the null hypothesis that there is no difference in bot score distributions and use Mann-Whitney U test to compare the distributions. We find statistically significant differences with the normal sample for each coordinated group.
As seen in Fig~\ref{fig:bot_scores}, for the suspicious Bioweapon (CCP) group the (z-score -2.56, p-val 0.00523) are significant at 0.05. Similarly for the suspicious Great Reset conspiracy, at 0.1 significance level, suggesting higher distribution of automated behaviours in the detected coordinated account groups.

\textbf{Fig~\ref{fig:infl_coordinated}}
examines what the model learns from observed account activities in detecting coordinated groups. We obtain the estimates of the \textbf{mutual triggering effect (or influence)} between accounts from the learned model. The model estimates the density of future activities on the network given past activities, encoding which account pairs trigger each other's activities. Fig~\ref{fig:infl_coordinated} shows the average influence weight from accounts in one group (Influencer) on accounts in other groups (Influenced). As we observe, the model picks up stronger influence within accounts of the suspicious coordinating groups. Weaker influence patterns with normal accounts indicate more random activities of normal accounts (that may not be centrally controlled, externally colluding, or jointly collaborating to promote agendas). Also, the model does not find hidden influence across accounts in the two coordinated groups, suggesting presence of separate efforts, as is also evident from the separate agendas of the two groups (Bioweapon (CCP) vs. Great Reset conspiracy, as seen in Fig~\ref{fig:coordinated_hashtags}).

\section{Misinformation and Information Communities}

In the vaccine discussion, earlier works have observed echo-chambers or communities of accounts with similar opinions that endorse and re-share (retweet) each other's content \cite{garimella2018quantifying,cossard2020falling}. Since retweets are a form of endorsement, we identify community structure of accounts from the retweet graph.

\subsection{Structure} Similar to prior work \cite{garimella2018quantifying}, we use RT edges with minimum count of retweets $>= 2$, including mutual retweets. We restrict the analysis to accounts appearing at least 5 times in the dataset, to ensure that the collected tweets sampled by Twitter API contain enough information about the account. For the RT graph, we use the 3-core decomposition to exclude users with only weak connections to the primary
discussions \cite{miyazaki2021strategy}. %of the graph, to restrict our focus to accounts that are a part of significant community structuring of the diffusion network and more prone to echo chamber effects from communities they associate with.
We split the tweets in the dataset into four quarters by the timeline of collected tweets, and construct the RT graph on the first quarter. The decomposition helps to limit the size of the network for easier inspection and visualization of the graph for characterization of the communities. % Need to inspect future communities and justify this finding
We obtain a graph with 91k accounts, 121k edges, and avg. degree 2.66 before the k-core decomposition. After 3-core decomposition we have 8,974 accounts with 31k edges and an avg. degree of 6.92. We applied \emph{Louvain} community detection \cite{blondel2008fast} and obtained 39 communities. % Without k-core decomposition, 4634 communities are obtained using Louvain.

% \begin{figure}
%     \centering
%     \includegraphics[width=1.1\columnwidth,height=5cm]{images/RT_communities.png}
%     \caption{Characterization of COVID-19 vaccines discussion communities in 3-core of RT graph.}
%     \label{fig:RT_coms}
% \end{figure}

\begin{table*}[t]
\centering
\resizebox{0.95\textwidth}{!}{
\begin{tabular}{p{1cm}|p{2cm}|p{3cm}|
p{1cm}p{1cm} p{1cm}|p{1.2cm} p{1.2cm}p{1.1cm}  |p{1cm}|p{3cm}|p{3cm}}
\toprule
 &   &  & \multicolumn{3}{|l|}{Inferred Political Leaning}  & \multicolumn{3}{|l|}{Low-Quality News URLs}     & & &  \\
 \midrule 
No. & Language  & Locations & Left & Right & Und & Conspiracy     & Unreliable     & Reliable & Others & Top News Domains  & Top Accounts \\
\midrule
0(20\%)                                        & EN (97.4\%)                                    & US (90.3\%)                                                                            & \textbf{98.7}                & 0.3                                                  & 1                                               & 0.26           & 2.87                                   & \textbf{96.87}                   & 92.22                                     & nytimes, washingtonpost, cnn, latimes, politico                        & JoeBiden, KamalaHarris, NYGovCuomo                \\
1(16\%)                  & EN (94.0\%)                                    & US (48.9\%), UK (27.5\%)                                                               & 2.8                          & 6.8                                                  & \textbf{90.4}            & \underline{\textbf{39.61}} & \underline{33.82}        & 26.57                            & 87.56                                     & childrenshealthdefense, dailymail, zerohedge, rt, lifesitenews         & ChildrensHD, zerohedge                            \\
2(16\%)                                       & EN (95.3\%)                                    & US (57.5\%), India (7.6\%), UK (6.1\%)                                 & \textbf{94.4}                & 1.2                                                  & 4.4                                                & 0.94           & 5.47                                   & \textbf{93.59}                   & 91.54                                     & reuters, theguardian, nytimes, independent, latimes                    & Reuters, NBCNews, AP, CoronaUpdateBot             \\
3(11\%)               & EN (96.9\%)                                    & US (86.6\%)                                                                            & 5.8                          & {\textbf{83.9}}                 & 10.3                                               & \underline{32.31}          & \underline{\textbf{35.14}} & 32.55                            & 89.62                                     & truepundit, foxnews, theepochtimes, zerohedge, dailymail               & Mike\_Pence, GOPChairwoman, OANN, nypost \\
4(6\%)                                        & EN (84.5\%)                                    & India (90.3\%)                                                                         & 8.2                          & 0.2                                                  & { \textbf{91.6}}               & 10.27          & 37.76                                  & \textbf{51.97}                   & 97.92                                     & swarajyamag, indianexpress, dailymail, nationalfile, wsj               & timesofindia, WIONews, mygovindia                 \\
5(5\%)                                        & EN (96.6\%)                                    & UK (76.9\%), US (13.8\%)                                                               & \textbf{64.6*}               & 0.2                                                  & 35.2                                               & 1.28                                   & 10.25                                  & \textbf{88.47}                   & 93.12                                     & theguardian, nytimes, telegraph, bbc, express                          & DHSCgovuk, NHSEngland, NHSuk                      \\
6(5\%)                                        & \textbf{ES (81.2\%), EN (13.3\%)}              & Argentina (34.3\%), Mexico (11.9\%)                                                    & \textbf{56.9*}               & 0                                                    & 43.1                                               & 3.48           & 7.32                                   & \textbf{89.2}                    & 97.68                                     & nytimes, reuters, bbc, theguardian, thetimes                           & ReutersLatam, CoronavirusNewv, AlertaNews24       \\
7(3\%)                                        & EN (95.6\%)                                    & Canada (91.0\%)                                                                        & \textbf{87.8}                & 0                                                    & 12.2                                               & 1.78           & 4.07                                   & \textbf{94.15}                   & 97.05                                     & nytimes, theguardian, reuters, washingtonpost, bloomberg               & JustinTrudeau, CBCAlerts, CPHO\_Canada            \\
8(3\%)                & EN (97.3\%)                                    & US (94.8\%)                                                                            & 0                            & { \textbf{97.7}}                 & 2.3                                                & \underline{31.86}          & \underline{\textbf{45.14}} & 23                               & 85.09                                     & thegatewaypundit, breitbart, foxnews, lifesitenews, zerohedge          &                                                   \\
9(3\%)               & \textbf{FR (81.9\%), EN (10.6\%)}              & France (92.0\%)                                                                        & 4                            & 0                                                    &  \textbf{96}                & \underline{14.18}          & \underline{\textbf{62.88}} & 22.94                            & 90.18                                     & francesoir, fr, reseauinternational, childrenshealthdefense, dailymail & sputnik\_fr, VirusWar, franceinfo, afpfr          \\
10(3\%)                                       & EN (90.8\%)                                    & India (84.2\%)                                                                         & \textbf{92.3}                & 0                                                    & 7.7                                                & 0              & 2.66                                   & \textbf{97.34}                   & 95.92                                     & nytimes, indianexpress, reuters, theguardian, bloomberg                & CNBCTV18News                                      \\
11(2\%)                                       & EN (79.2\%), TL (18.8\%)                       & Philippines (85.7\%)                                                                   & \textbf{100}                 & 0                                                    & 0                                                  & 7.5            & 1.25                                   & \textbf{91.25}                   & 98.25                                     & reuters, buzzfeednews, theguardian, nytimes, prevention                & ANCAlerts, CNNPhilippines                         \\
12(2\%)               & \textbf{EN (53.5\%), IT (38.7\%)}              & Italy (32.7\%), US (20.4\%)                                                            & 34.5                         & 1.9                                                  & { \textbf{63.6*}}              & \underline{22.22}          & \underline{\textbf{55.19}} & 22.59                            & 91.28                                     & imolaoggi, zerohedge, rt, dailymail, nytimes                           & RT\_com, SputnikInt                               \\
13(2\%)                                       & EN (88.4\%)                                    & US (30.8\%), South Africa (13.8\%)                                                     & \textbf{86.9}                & 0.8                                                  & 12.3                                               & 1.79           & 1.57                                   & \textbf{96.64}                   & 90.33                                     & npr, nytimes, latimes, theguardian, wsj                                & NPRHealth, WHO, EU\_Health, CovidSupportSA        \\
14(1\%)               & \textbf{ES (48.6\%), EN (26.0\%), NL (18.9\%)} & Netherlands (28.6\%), Uruguay (14.3\%), Latvia (14.3\%) & \textbf{60.3*}               & 1.8                                                  & 37.9                                               & \underline{\textbf{49.52}} & \underline{29.61}          & 20.87                            & 90.84                                     & humansarefree, dailymail, rt, childrenshealthdefense, zerohedge        &    \\                    
\bottomrule
\end{tabular}
}
\caption{Misinformation and information communities in COVID-19 vaccine 3-core of retweet graph. Misinformation communities are highlighted based on proportion of unreliable/conspiracy tweets and correspond to anti-vaccine/far-right groups.}
\label{tab:coms}
\end{table*}

\subsection{Characterization} We characterize the top-20 diffusion communities that account for 96\% of the accounts in the 3-core RT graph. Fig.~\ref{fig:RT_coms} presents the communities with its characterization in Table.~\ref{tab:coms} based on nature of accounts in tweets in each community. The table includes the community number with size (\% accounts) in the graph, language in tweets from the community, geolocation extracted from geo-enabled tweets/reported valid locations in account profiles \cite{dredze2013carmen} to characterize the general demographic of the community. In addition, we infer the political leaning of accounts using left/right media URLs (as classified by allsides.com) endorsed directly or through retweet structure, similar to \cite{badawy2019characterizing}. We jointly inspect these with  the top retweeted accounts and tweets, and distribution of URL news sources, top URL/news domains, and contents in top retweeted and random subset of tweets. %The main findings indicate,
\begin{itemize}
    \item A large (16.24\%) community (C1) of Anti-vaccine misinformation and conspiracies. From accounts that have valid locations reported in the profile or tweets, this community spans US (48.9\%) and UK (27.5\%).
    \item Other smaller communities with dominant misinformation or conspiracy tweets correspond to the U.S. Far-right conspiracy group that post anti-vaccine content (C8). Another Spanish-English tweets community (C14) contains strongly anti-vaccine conspiracies, very similar to the larger Anti-vaccine misinformation community. A French tweets community (C9) with relatively less conspiracy content but anti-vaccine, and an Italian tweets community (C12) of mixed stance to vaccine hesitancy are present close to the anti-vaccine community.
    \item Benign communities included Mainstream News (15.58\%) (C2), Health news (1.48\%) (C13), U.S. Left Leaning (C0) (with Joe Biden, Kamala Harris as top retweeted accounts). The former contain accounts with more global geolocations, while the latter was dominantly with US geolocations. There are several regional news and politics communities centered close to the Mainstream and Health News communities (e.g. UK based with top retweeted accounts corresponding to the National Health Service NHS and Department of Health and Social Care DHSC (C5), Latin America (C6), Philippines News (C11), India and Canada News and Politics (C4, C10 and C7)). These are identified based on tweets language, contents and inferred geolocations.
    \item The U.S. Right-leaning community (C3) (Mike Pence, OANN, top Republicans as most retweeted accounts), however different from other communities, has roughly equal proportions of unreliable, conspiracy and reliable URLs in their tweets. In terms of tweet contents and proximity to other communities, the right-leaning community is closer to the Anti-vaccine misinformation and conspiracy community (C1), as well as the far right group (C8), with relatively sparse edges to the global mainstream informational community (C2,C13). The top news domains in their tweets include relatively credible sources with possible right-leaning bias such as Foxnews, and also conspiracy sources like zero-hedge and Truepundit. Inspecting top retweeted and random sample of tweets suggests a mixture of vaccine news updates, politically biased views and also conspiracies (e.g. numerous anti-China messages, Bill Gates conspiracies), and largely anti-vaccine/protocol stance (including misinformation about the vaccines). The misinformation and information communities are strongly separated with sparse RT edges between them (as seen from the separation in the RT graph visualization), the proximity within different anti-vaccine misinformation communities, and within pro-vaccine/news dense communities being higher.
\end{itemize}

In additional details of the table, note that unreliable, conspiracy, reliable news source URLs proportions are included. For tweets containing no URLs or unidentified URL domains, the fraction of such tweets is listed under Others column in the Table.
The inferred political leanings that are for majority of the accounts in a community are highlighted in bold or color. The asterisk is used for communities with more than a single prominent inferred leaning type. Languages and geolocations smaller than 10\% in the tweets are not reported. Top News Domains and Top Accounts are presented examples from the top-20 in each community, and left unmentioned if the top-20 contained only individuals that were not agencies or well known figures (e.g. C8). % , example in the U.S. Far-right conspiracy community (8).   

\subsection{Feature Distributions} 
\begin{figure*}[t]
    \centering
    \begin{subfigure}{0.32\textwidth}
    \centering
    \includegraphics[width=\textwidth,height=2.5cm]{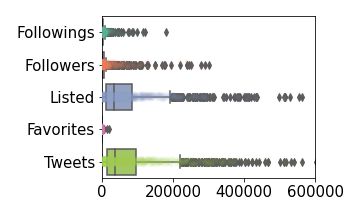}
    \caption{Anti-vax misinformation/ conspiracy}
    % \label{}
    \end{subfigure}
    ~
    \begin{subfigure}{0.33\textwidth}
    \centering
    \includegraphics[width=\textwidth,height=2.5cm]{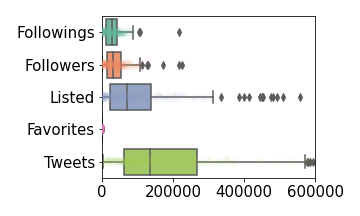}
    \caption{Far-right leaning conspiracies}
    % \label{}
    \end{subfigure}
    ~
    \begin{subfigure}{0.32\textwidth}
    \centering
    \includegraphics[width=\textwidth,height=2.5cm]{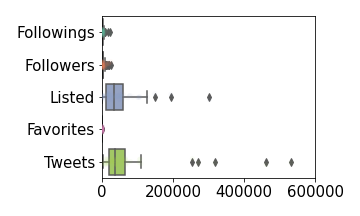}
    \caption{Spanish-En anti-vax conspiracies}
    % \label{}
    \end{subfigure}
    ~
    \begin{subfigure}{0.33\textwidth}
    \centering
    \includegraphics[width=\textwidth,height=2.5cm]{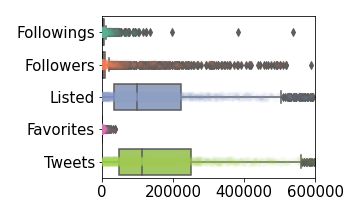}
    \caption{Left-Leaning}
    % \label{}
    \end{subfigure}
    ~
    \begin{subfigure}{0.32\textwidth}
    \centering
    \includegraphics[width=\textwidth,height=2.5cm]{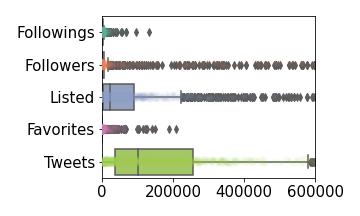}
    \caption{Mainstream News}
    % \label{}
    \end{subfigure}
    ~
    \begin{subfigure}{0.32\textwidth}
    \centering
    \includegraphics[width=\textwidth,height=2.5cm]{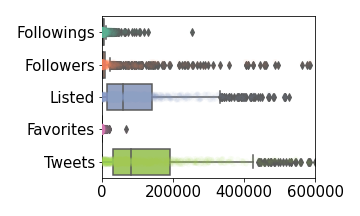}
    \caption{Right-leaning}
    % \label{}
    \end{subfigure}
    \caption{Account statistics boxplot for information and misinformation / conspiracy communities.}
    \label{fig:com_user_stats}
\end{figure*}

\begin{figure*}[t]
    \centering
    \begin{subfigure}{0.32\textwidth}
    \centering
    \includegraphics[width=\textwidth,height=3cm]{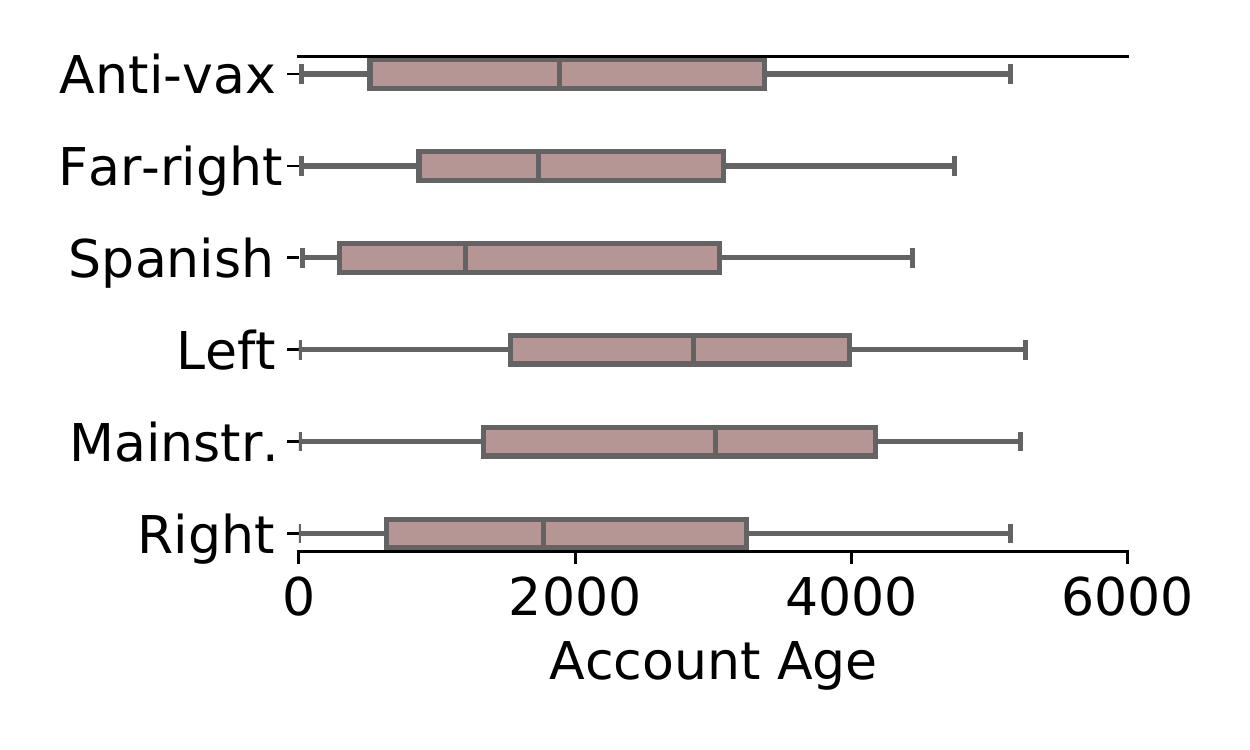}
    % \caption{Far-right leaning conspiracies}
    % \label{}
    \end{subfigure}
    ~
    \begin{subfigure}{0.32\textwidth}
    \centering
    \includegraphics[width=\textwidth,height=3cm]{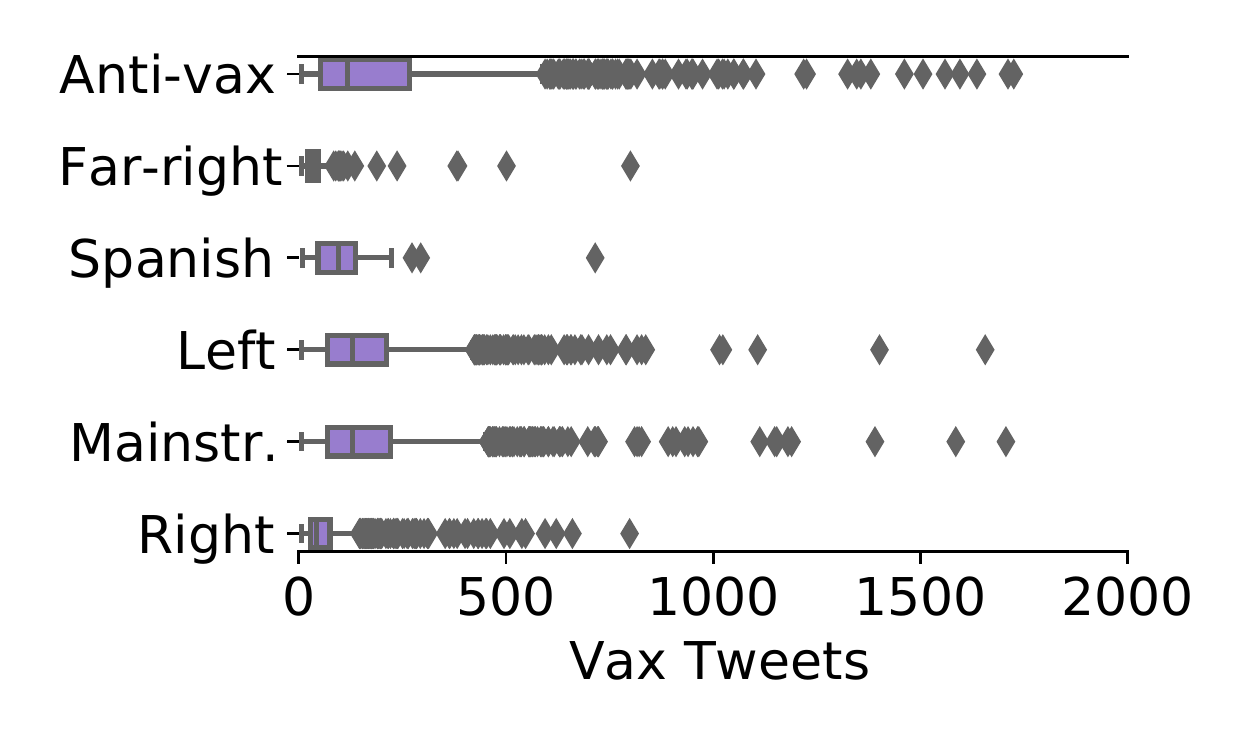}
    % \caption{Spanish-En anti-vax conspiracies}
    % \label{}
    \end{subfigure}
    ~
    \begin{subfigure}{0.32\textwidth}
    \centering
    \includegraphics[width=\textwidth,height=3cm]{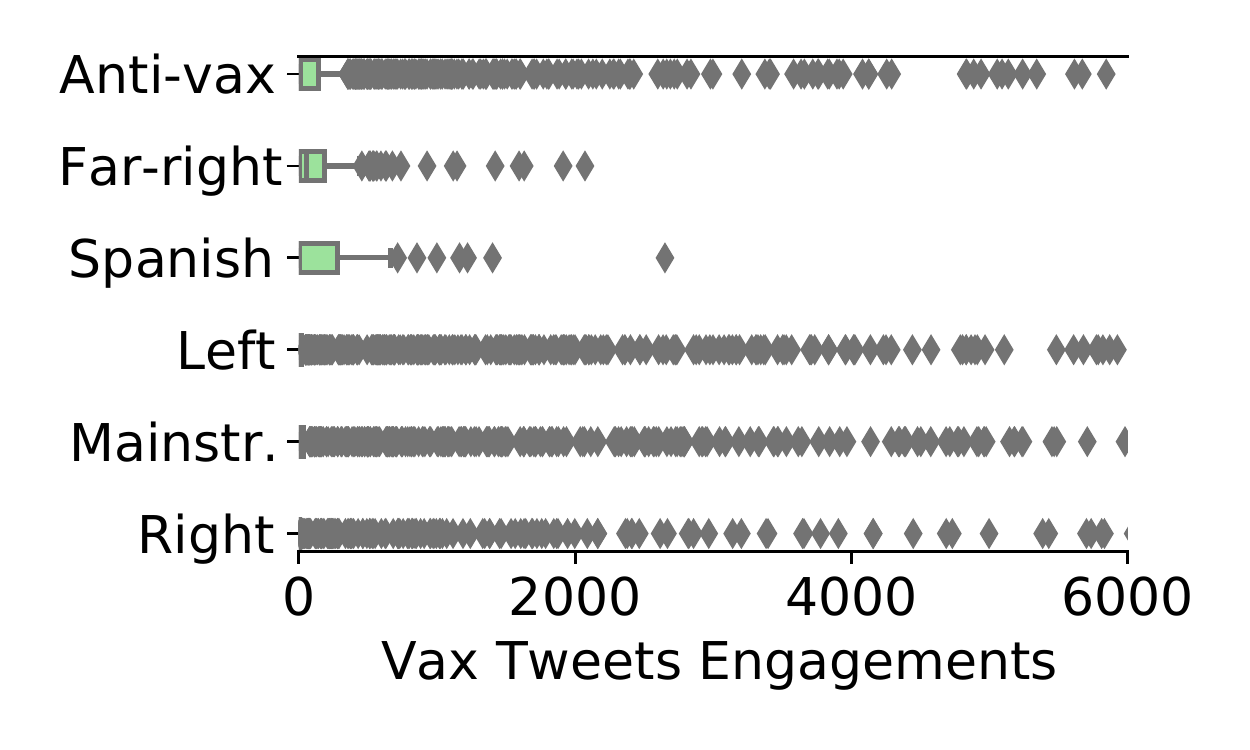}
    % \caption{Left-Leaning}
    % \label{}
    \end{subfigure}
    \caption{Account statistics for communities (a) account age, (b) number of collected vaccine tweets (original, retweet, reply) by account, and Engagements (account retweeted, replied, mentioned by others) in collected vaccine tweets.}
    \label{fig:com_user_stats_more}
\end{figure*}

The feature distributions of accounts in the most significant communities (circled in Fig~\ref{fig:RT_coms}) are compared in Fig.~\ref{fig:com_user_stats} and Fig.~\ref{fig:com_user_stats_more}. In Fig.~\ref{fig:com_user_stats}, we compare Followings (accounts followed by accounts in the community), Followers, Listed (accounts mentioned in topical lists created by other accounts on Twitter), Favourites (number of tweets liked/favourited by the accounts in the community), Tweets (which is the total tweets posted by the account in its lifetime). These features are available in the account metadata obtained using the Twitter API, and we use the account's statistics at its last observed tweet in the dataset. The main insights about account characteristics in each community are,
\begin{itemize}
    \item Followers and Followings distribution interquartile range is significantly higher for the Far-right conspiracy community (C8) (Fig.~\ref{fig:com_user_stats}). The distribution across other groups is similar, with the Anti-vaccine community (C1) having slightly lower upper quartile % (more fine-grained plots are in the Appx. Fig~\ref{fig:com_user_stats_appx}). 
    This suggests more interconnected accounts in the Far-right conspiracy group, which likely actively follow each other.
    \item More accounts appear in Twitter Lists (Listed) from Left (C0), Mainstream (C2), Right (C3), and Far-right communities, compared to Anti-vaccine misinformation and Spanish-En conspiracy community (C14). Yet, the anti-vaccine community do have many Listed accounts, suggesting that such content does have a sizable audience.
\end{itemize}

In Fig~\ref{fig:com_user_stats} we also have the Tweets (total tweets posted by the account in its lifetime). To better understand the accounts activities, in Fig~\ref{fig:com_user_stats_more}, we additionally compare the Account Age (Days between first observed tweet in the dataset and the account creation date), Vax Tweets (tweets posted by the account specific to vaccine related content, quantified by observed tweets of the account in the collected dataset), Vax Tweet Engagements (number of vaccine tweets that reference i.e. mention, reply, retweet, quote the account in the community or any of its vaccine tweets, quantified by observed tweets in the collected dataset). The findings are,
\begin{itemize}
    \item Total Tweets distribution is lowest for Anti-vaccine misinformation and Spanish-En conspiracy communities.  
    \item Account Age distribution is the smallest for these two communities, i.e., more recently created accounts. Account Age for Left and Mainstream are highest (older accounts), and Right and Far-right are in between.
    \item In the Vax Tweets, however, the Anti-vaccine community is the most vocal with higher Vax Tweet distribution than other communities; Far-right and Right being the least.
    \item Interestingly, Anti-vaccine misinformation, Far-right conspiracy, and Spanish-En conspiracy communities have the largest interquartile ranges compared to the other communities on Vax Tweets Engagements. That means most accounts in these communities receive non-negligible engagements (mentions from other accounts, or replies, retweets, quotes of its vaccine tweets), in contrast with the other communities. It suggests that discussions are lead in a more decentralized manner in these communities.
\end{itemize}

\subsection{Vaccine Uptake Distribution in U.S. States}

In Fig.~\ref{fig:geo_dist_misinfo}, we compare the ratio of misinformation (unreliable/conspiracy URLs) to reliable URLs observed in collected tweets for accounts with geolocation available extracted using \cite{dredze2013carmen}. The public records of how many people have been vaccinated in each state available through CDC is curated and maintained by the research community \cite{mathieu2021global} (accessed June 6, 2021). 

\subsubsection{Analysis.} 
\begin{figure}[t]
    \centering
    \includegraphics[width=\columnwidth,height=5cm]{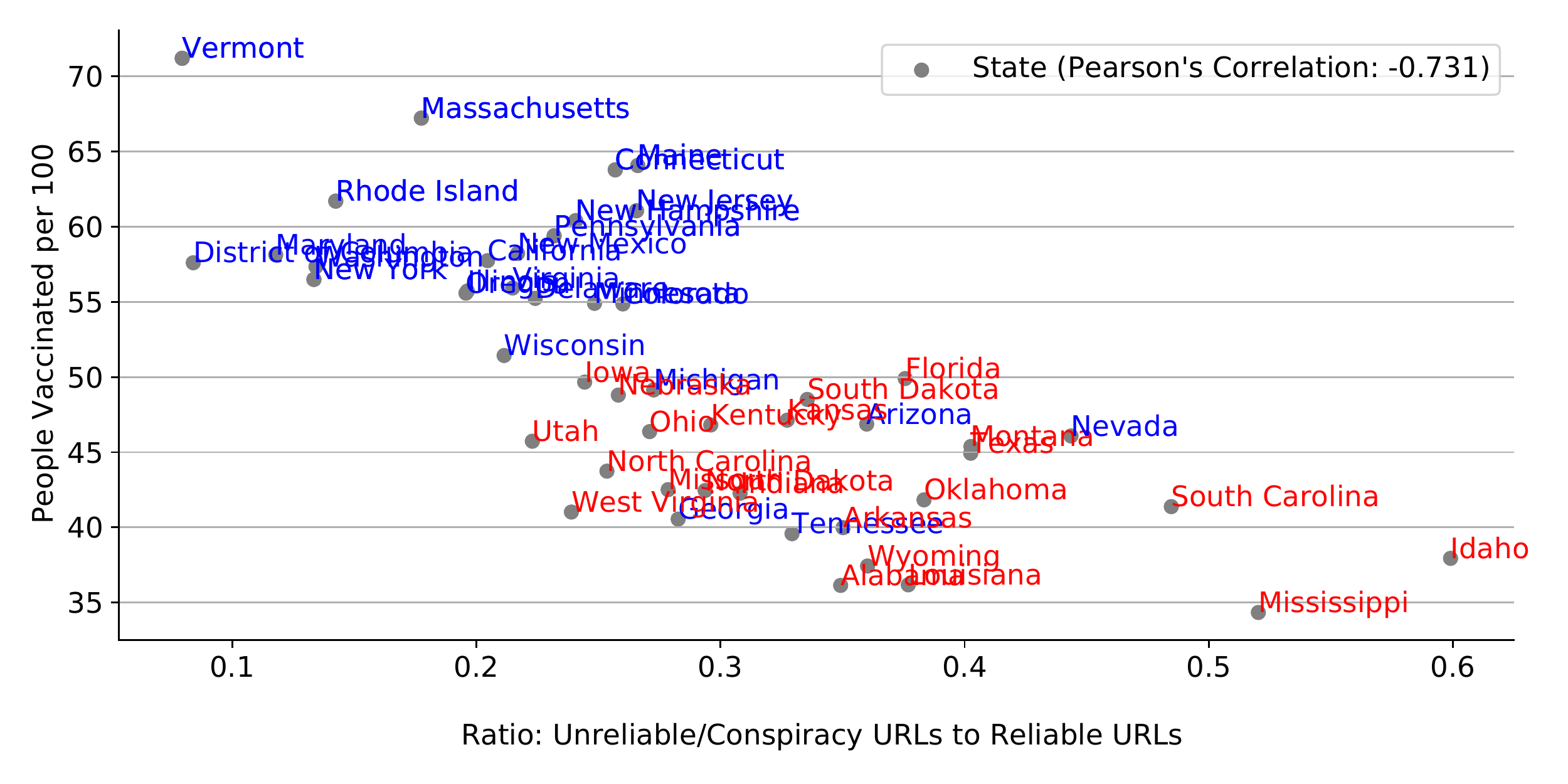}
    \caption{\% Vaccinated individuals  in each U.S. states vs. ratio of unreliable/conspiracy to reliable URL tweets (Pearson coefficient: -0.731, p-val: 7.5e-11) extracted with geolocation over Red and Blue states in 2020 Elections. }
    \label{fig:geo_dist_misinfo}
\end{figure}

The vaccine uptake i.e., percentage of vaccinated individuals as of June 6, 2021 per U.S. state is plotted against the rate of misinformation (ratio of unreliable/conspiracy URL tweets to reliable URL tweets) in Fig~\ref{fig:geo_dist_misinfo}. The states political affiliation is designated based on the 2020 Election votes (Red states voted for Donald Trump and Blue States for Joe Biden in the 2020 Presidential Election). The analysis is state-wise, therefore, only account tweets with valid extracted geolocations of US States are utilized in the plot. The estimated correlation results are,
\begin{itemize}
    \item The Pearson's correlation coefficient is -0.731 between vaccine uptake and misinformation rate. This confirms a high negative correlation of \% individuals vaccinated and the rate of online misinformation across states.
    \item The vaccination uptake is consistently lower in Republican (red or right-leaning) states, and some swing states e.g. Nevada, Arizona; while the rate of misinformation is consistently higher. This politically biased segregation of vaccine hesitancy or anti-vaccine sentiments is also evident from the misinformation diffusion community analysis (discussed in the prior subsections).
\end{itemize}

% \begin{figure*}[t]
%     \centering
%     \begin{subfigure}{0.48\textwidth}
%     \centering
%     \includegraphics[width=\textwidth,height=4cm]{images/us_states_misinfo_prop_to_rel_tweets_till_Apr24.png}
%     \caption{Ratio of unreliable/conspiracy to reliable URL tweets with extracted geolocation over U.S. states.}
%     % \label{}
%     \end{subfigure}
%     ~
%     \begin{subfigure}{0.48\textwidth}
%     \centering
%     \includegraphics[width=\textwidth,height=5cm]{images/us_states_vaccine_uptake.pdf}
%     \caption{\% Vaccinated individuals  in each U.S. states vs. ratio of unreliable/conspiracy to reliable URL tweets extracted with geolocation over Red and Blue states in 2020 Elections.}
%     % \label{}
%     \end{subfigure}
%     \caption{Unreliable/conspiracy URLs and Reliable URLs distribution based on extracted geolocation from tweets.}
%     \label{fig:geo_dist_misinfo}
% \end{figure*}

\section{Misinformation Narratives of Vaccines}

Social media communication studies have outlined factors that can mislead the public and increase vaccine hesitancy \cite{lewandowsky2021covid}. The five techniques of science denial are provided under the acronym FLICC (fake experts, logical fallacies, impossible expectations, cherry-picking, and conspiracy theories). In this section, we look at different types of narratives and contexts of misinformation. % through which vaccine hesitancy on social media can be increased.

\subsection{Social Media Discussion of Vaccination Effects}

We examine vaccine side-effects discussed on social media in misinformation narratives. We study whether the discussion of vaccine side-effects on social media differs from the CDC VAERS (accessed June 10, 2021) recorded side-effects obtained from healthcare providers and public reports. 

% Besides misinformation and conspiracy narratives that can decline vaccination intentions \cite{jolley2014effects}
% The biased discussion of vaccine side-effects or adverse effects % post vaccination can again promote hesitancy. This 
% falls under cherry-picking, one of the five science denial techniques. % Here, we study whether the discussion of vaccine side-effects on social media differs from the CDC VAERS recorded side-effects obtained from healthcare providers and public reports. %, and if the differences might have potential to increase COVID-19 vaccine hesitancy.

% Using the CDC VAERS records accessed June 10, 2021, we hope to answer the following questions,
\begin{enumerate}
    \item Are the side-effects widely discussed in vaccine related tweets also common in VAERS reports?
    \item Is the distribution of side-effects discussed  in misinformation URL tweets distinct from all tweets?
\end{enumerate}

To answer the above questions, we explore the correlation between frequency of the discussed side-effects on social media and their frequency in the VAERS records. To measure the frequency on Twitter, we first extract the medical concepts from the tweets via text matching based on a large medical concept corpus: AskAPatient \cite{limsopatham2016normalising}. This corpus provides us with common medical concepts on social media in different forms, such as abbreviations, complete names and even misspelling versions. We use the number of tweets in which a medical concept appears to represent its frequency on Twitter. To measure the frequency in VAERS records, we conduct medical concept extraction in the same way and count the number of individuals whose medical records mention a concept.

\subsubsection{Analysis.} 
\begin{figure*}[t]
    \centering
    \begin{subfigure}{0.4\textwidth}
    \centering
    \includegraphics[width=\columnwidth,height=5cm]{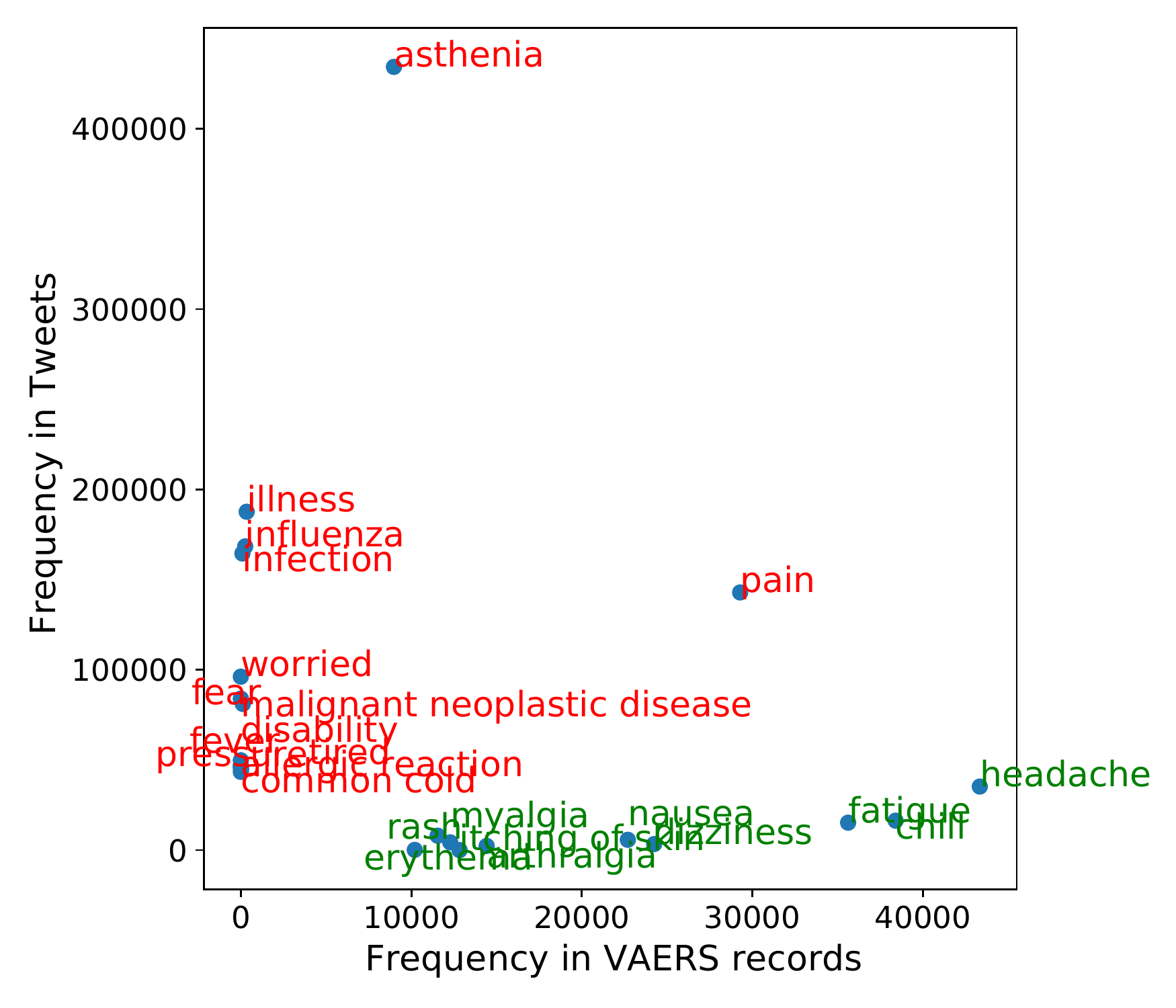}  % tweet_set.png
    \caption{All tweets in dataset. (Pearson coeff. -0.250, p-val 0.239). Tweet frequency vs. VAERS records frequency for reported medical concepts post vaccination.}
    \label{fig:vaers_all_tweets}
    \end{subfigure}
    ~
    \begin{subfigure}{0.4\textwidth}
    \centering
    \includegraphics[width=\columnwidth,height=5cm]{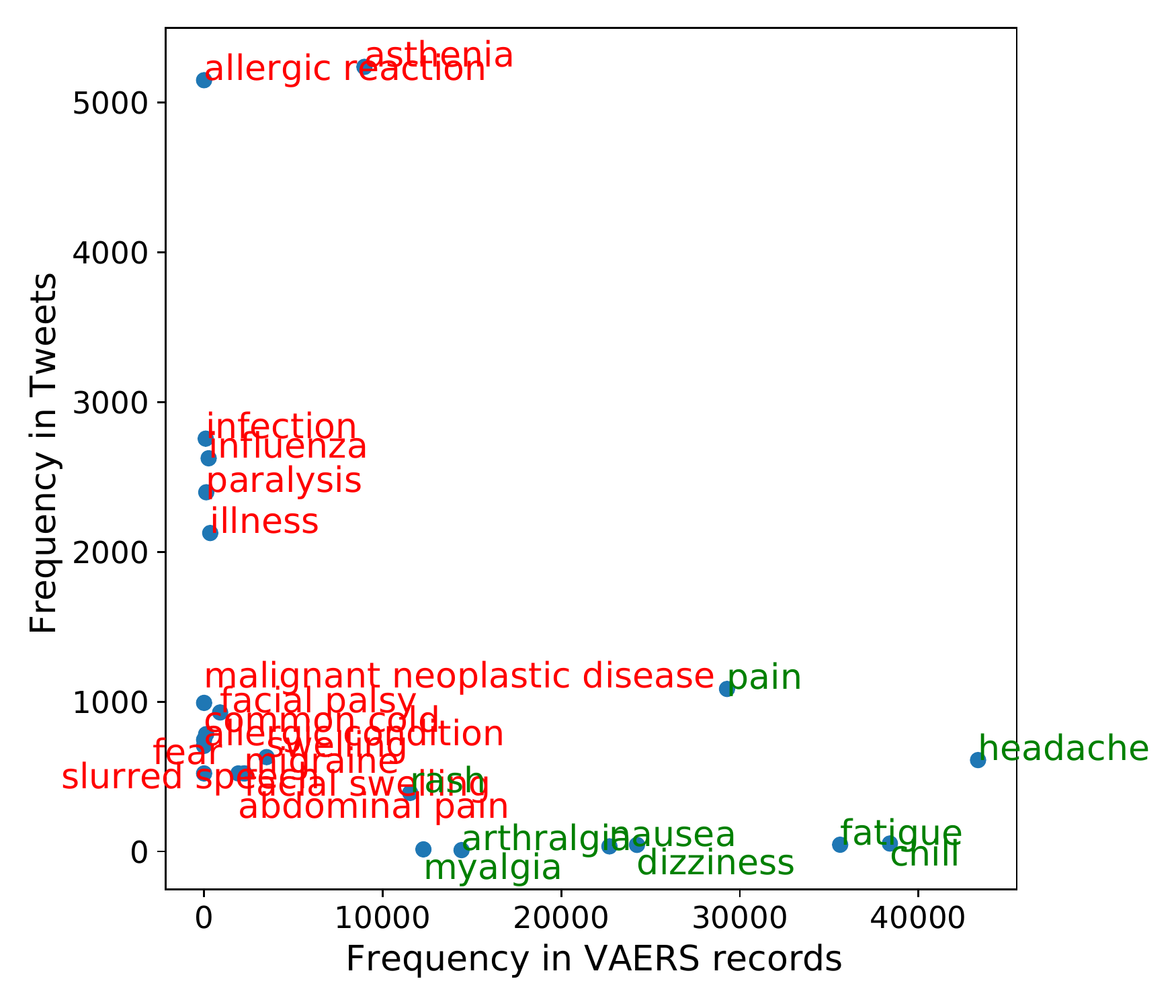} % ns_tweet_set.png
    \caption{Unreliable/conspiracy URLs tweets. (Pearson coeff. -0.358, p-val 0.079). Tweet vs. VAERS record frequency for reported medical concepts post vaccination.}
    \label{fig:vaers_misinformation_url_tweets}
    \end{subfigure}
    \caption{Frequency correlation of side-effects discussed on Twitter compared with that recorded in VAERS.}
    \label{fig:coorelation_vaers}
\end{figure*}

In Fig.~\ref{fig:coorelation_vaers}, we plot the medical concept frequencies in all collected tweets (Fig.~\ref{fig:vaers_all_tweets}) and in unreliable/conspiracy URLs tweets (Fig.~\ref{fig:vaers_misinformation_url_tweets}) against corresponding frequencies in VAERS records. From the visualization, we can see that the concepts that are widely discussed on social media are the ones that are rarer in the medical records. More frequent effects like pain, fever, headache are less frequent in tweets. In misinformation URL tweets, rarer reported concepts such as ``paralysis", ``allergic reaction", ``malignant neoplastic disease" (cancer) are more frequently referenced. The biased discussion of vaccine side-effects or adverse effects falls under cherry-picking, one of the five science denial techniques. This nuanced distortion of the true facts for misleading narratives is especially challenging in COVID-19 vaccine misinformation and harder to mitigate.

% The negative correlation between discussed concepts and VAERS reports increases the potential for vaccine hesitancy in social media users, since rare symptoms get more attention due to its novelty in all tweets or due to cherry-picking in unreliable/conspiracy tweets. The Pearson correlation coefficient is more negative for unreliable/conspiracy URL tweets at -0.36 compared to -0.25 on all tweets.

% \begin{wraptable}{r}{0.4\textwidth}
% % \begin{table}[t]
% % \centering
% \caption{Pearson Coefficient between the Twitter Frequency and Reported Frequency}
% \label{tab:Correlation}
%  \begin{tabular}{l|r}
% \toprule
% \textbf{Data Fraction} & \textbf{Pearson Coefficient}  \\
% \midrule
% All tweets & 0.4395 \\
% Misinformation tweets & 0.1345 \%  \\
% \bottomrule
% \end{tabular}
% % \end{table}
% \end{wraptable}

\begin{figure}[t]
    \centering
    \includegraphics[scale=0.35]{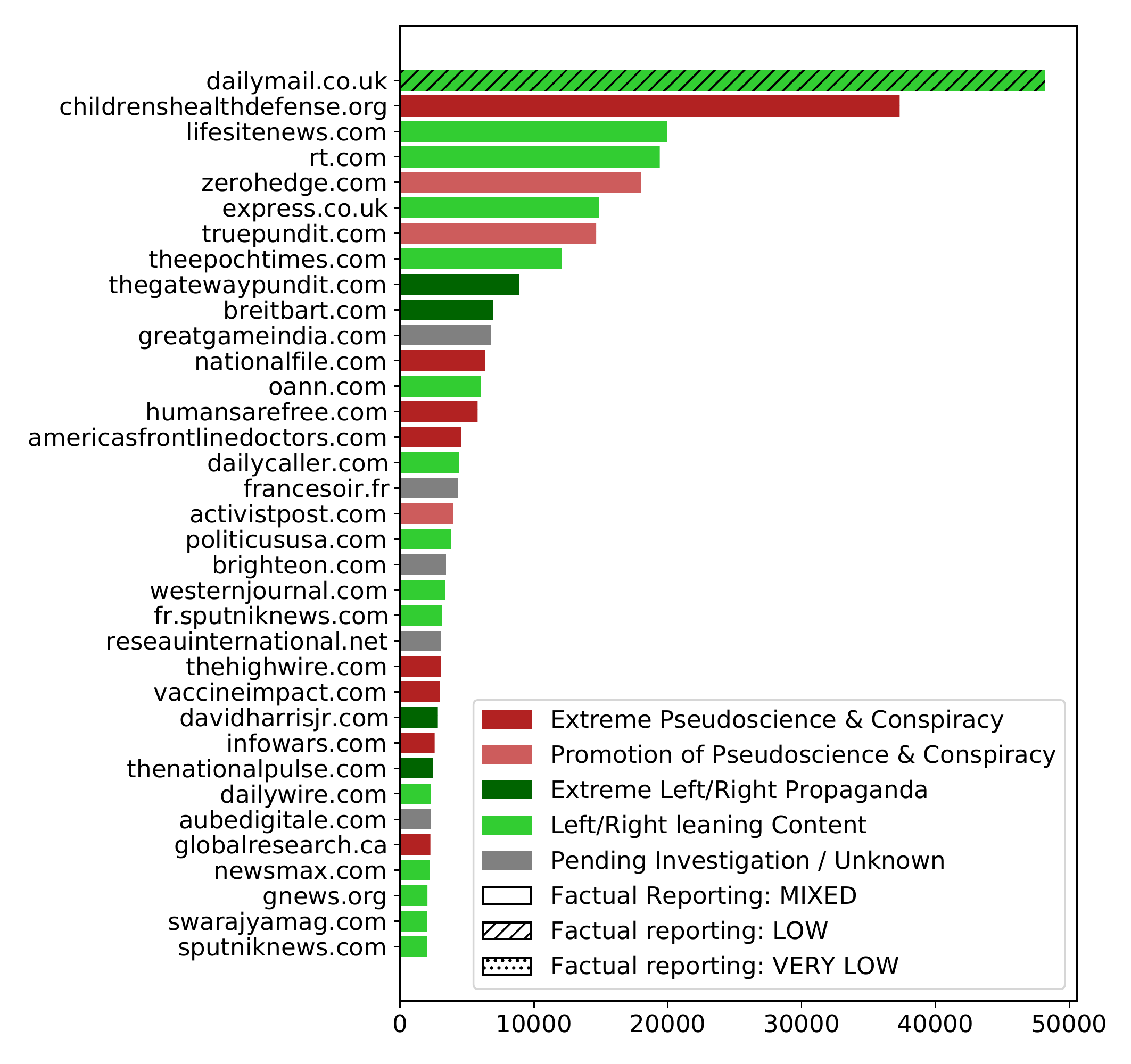}
    \caption{Tweet volume for top-35 news sources.}
    \label{fig:misinfo_news_source_categorization}
\end{figure}

% \begin{figure*}[t]
%     \centering
%     \includegraphics[width=0.8\textwidth,height=7.5cm]{images/zoom.pdf}
%     \caption{Comparison of symptom frequency on Twitter and official medical records of patients after vaccination: Visualization of all tweets. (Pearson correlation = 0.4395)}
%     \label{fig:all_all_plot}
% \end{figure*}

% \begin{figure*}[t]
%     \centering
%     \includegraphics[width=0.8\textwidth,height=7.5cm]{images/zoom_ns.pdf}
%     \caption{Comparison of symptom frequency on Twitter and official medical records of patients after vaccination: Visualization of misinformation tweets. (Pearson correlation = 0.1345)}
%     \label{fig:ns_all_plot}
% \end{figure*}

\subsection{Topic Modeling of Misinformation Tweets}

We use topic modeling on unreliable/conspiracy URLs tweets. The text is pre-processed by tokenization, punctuation removal, stop-word removal, and removal of URLs, hashtags, mentions, and special characters, and represented using pre-trained \texttt{fastText} word embeddings \cite{bojanowski2017enriching}\footnote{Pre-trained embeddings can be downloaded from https://fasttext.cc/docs/en/english-vectors.html}. The average of word embeddings in the tweet text is used to represent each tweet. Pre-trained embeddings trained on large English corpora encode more semantic information useful for short texts where word co-occurrence statistics are limited for utilizing traditional probabilistic topic models \cite{li2016topic}. The text representations are clustered (k-means) to identify topics. Number of clusters is selected using silhouette and Davies-Bouldin measures of cluster separability between 3-35. Inspecting the word distribution and top representative tweets, we label and merge over-partitioned clusters.

\begin{table*}[t]
    \centering
    \resizebox{17cm}{!} {    \begin{tabular}{p{2cm}|p{8cm}|p{11cm}}
        \toprule 
        Cluster Index & Top (tf-idf) words & Representative Tweets (nearest distance to centroid) \\
        \midrule 
        1 Scientific facts & mrna, pfizer, cells, moderna, human, via, system, new, operating, evidence, experimental, pathogenic, trial, priming, dna, virus, alarming, adults, analysis, study, hiv, correlation, older, fetal, flu & ``@naomirwolf @naomirwolf Zuckerberg refuses Covid vaccine due to real possibility of permenent DNA alteration. Zuckerbefg:I Share Some Caution on this [Vaccine] Because We Jus Don't Know the Long-Term Side Effects of Basically Modifying People's DNA and RNA https://t.co/R2MS3LexeR" \newline ``\#Nuremberg \#CrimesAgainstHumanity \#Plandemic \#CovidHOAX \#GreatReset \#Event201 \#Agenda2030 \#NoMasks \#NoForcedVaccines \#Pharmageddon   Mainstream science admits COVID-19 vaccines contain mRNA “nanoparticles” that trigger severe allergic reactions  https://t.co/KAYqrHRK8Y" \\
        % ``Vaccine rollout will trigger new Covid variants, Oxford scientist warns, adding ‘new layer of complexity’ to pandemic fight. Just how much more evidence do we need? The fake `vaccine` is part of the GATES controlled depopulation programme https://t.co/176N1Lk5Hl" \newline
        
        2 Side Effects & allergic, reaction, pfizer, fda, severe, adverse, serious, worker, health, healthcare, hospitalized, effects, moderna, news, workers, side, threatening, intubated, rate, doctor, people, boston, alaska, facial, higher, shot, suffering, paralysis, uk & ``Cardiothoracic surgeon warns FDA, Pfizer on immunological danger of COVID vaccines in recently convalescent and asymptomatic carriers | Opinion | LifeSite https://t.co/yPn3rSYxjy" \newline
        ``SAFE AND EFFECTIVE! IS IT? TIME TO FACE THE TRUTH!  FDA Investigates Allergic Reactions to Pfizer COVID Vaccine After More Healthcare Workers Hospitalized • Children's Health Defense https://t.co/HHfv3sbfGF" \\
        
        3 Effectiveness & news, pfizer, dr, says, world, via, fauci, bill, gates, kennedy, new, jr, robert, moderna, take, people, get, doctors, uk, coronavirus, rollout, desantis, video, taking, trial, pharma, african, us, medical, lifesite, ron, tests, gov, warns, big & 
        ``Can this this be?  Then Its Not a Vaccine: Crazy Dr. Fauci Said in October Early COVID Vaccines Will Only Prevent Symptoms and NOT Block the Infection What?  \#FREESPEECH \#WALKAWAY \#DEPLORABLE \#DrainTheSwamp \#FakeNews \#Trump2020 \#Israel \#StopTheSteal   https://t.co/4l9yGMIKM6" \newline ``This also happened with many vaccines, especially the polio. Hundreds of Israelis get infected with Covid-19 after receiving Pfizer/BioNTech vaccine – reports — RT World News https://t.co/5Swk5C5UBI" \\
        
        4 Deaths & pfizer, dies, receiving, home, nurse, nursing, days, getting, health, doctor, portuguese, worker, die, shot, taking, residents, first, old, two, miami, died, weeks, healthy, elderly  & ``46 Nursing Home Residents in Spain Die Within 1 Month of Getting Pfizer COVID Vaccine! @ScottMorrisonMP @GregHuntMP    Are you still proceeding with the \#Pfizer \#vaccine rollout which could kill older people?  \#BREAKING  \#BreakingNews \#auspol  \#COVID19   https://t.co/32CppATPpz" \\
        
        5 Vaccine Refusal & workers, pfizer, health, refuse, emergency, coronavirus, people, care, refusing, get, getting, use, hundreds, take, fda, staff, us, uk, room, says, passports, receiving, hospital, news, line, healthcare, one, report, doses
        & ``As many as 60\% of healthcare workers are refusing to get the \#Covid vaccine. There’s an overwhelming lack of trust. Why? https://t.co/5OCa5X5AHJ \#COVID19 \#vaccine \#CovidVaccine" \newline 
        ``Start a DEMONSTRATION AND BOYCOTT CAMPAIGN against the VACCINE TYRANTS! COVID vaccines disaster of Adverse reports to CDC....look it up!  NYC Waitress Fired For Refusing COVID Vaccine Over Fertility Concerns https://t.co/4vKDS8QjHw"
        \\
        
        6 Rollout & trump, biden, joe, uns, fetus, praises, gavi, playing, owns, million, funding, male, funded, gave, gates, stopped, billion, us, plan, admin, distribution, rollout, administration, google, jill, president, speed, doses, warp & ``Joe Biden Struggles to Read Teleprompter as He Trashes Trump Administration's Covid-19 Vaccine Distribution Efforts https://t.co/C1YOoU7XXH @gatewaypundit" \newline ``We all knew all along that Trump would botch the initial vaccine rollout, and that vaccinations wouldn’t properly get underway until Biden takes office. Just one more thing for Trump to screw up on his way down. https://t.co/V9jMVdcx04" \\
        % 7 Spanish tweets & & \\
        % 8 Italian tweets & & \\
        % 9 German tweets & & \\
        7 Dehumanization & takedowntheccp, yanlimeng, wipe, drlimengyan, weapon, bio, white, ccpvirus, war, made, plan, part, ccp, world & ``@Newsweek World War V5.0: Covid virus is a bio weapon made by CCP.  Vaccine is a part of plan to wipe out all white people. @DrLiMengYAN1  \#DrLiMengYan1  \#YanLiMeng  \#CCPVirus \#Covid19 \#TakeDownTheCCP %#闫丽梦 #爆料革命 #郭文贵  
        \#COVID19 https://t.co/i6P9uTe11F. https://t.co/4PYB8dxe2v https://t.co/j3oKSTlpa1" \\
        \bottomrule
    \end{tabular}
    }
    \caption{Misinformation topic clusters along with representative tweets and word distribution with highest TF-IDF scores.}
    \label{tab:misinfo_topics}
\end{table*}

\subsubsection{Analysis.} The misinformation largely targeted seven forms of information manipulation about the COVID-19 vaccines, namely, manipulation of Scientific facts about the COVID-19 vaccines, misleading information about Side Effects, Deaths, Vaccine Effectiveness, and Vaccine Refusal, along with misinformation related to Vaccine Rollout, and Dehumanization/Depopulation/Great Reset/Bill Gates/Pharmageddon conspiracies. Table.~\ref{tab:misinfo_topics} provides examples of top representative tweets and word distributions for each identified topic. Examination of the tweets suggests presence of the five techniques of manipulation FLICC \cite{lewandowsky2021covid} mentioned earlier. In part for vaccine safety and effectiveness, out-wright false claims about scientific facts, and side-effects existed, but also true reported side-effects were discussed with negative strong anti-vaccine sentiment, or missing or misleading contexts, including misleading expectations about vaccine effectiveness, by suggesting that since vaccines cannot prevent the infection, then its ineffective or not useful, as seen in Table.

\subsubsection{Frequent News Sources Categorization.} 

Fig.~\ref{fig:misinfo_news_source_categorization} presents categorization of the top news domains in unreliable/conspiracy URLs tweets. The categorization is based on the Media Bias/Fact Check ratings of (i) degree of factual reporting (ii) political bias, and (iii) scientific reporting measures. The factual reporting level from these sources is regarded as either Low, very Low, or Mixed on Media Bias/Fact Check. The top domains contain sources promoting both extreme pseudoscience and conspiracy (e.g. ChildrensHealthDefense.org), Left/Right political propaganda (e.g. dailymail.co.uk, rt.com, truepundit.com).  % In the news domain curation, we collected lists with Low or Very Low rating from MBF, but the mixed sources can appear if they are present in NewsGuard or Zimdars.
% \begin{wrapfigure}{r}{0.53\textwidth}
% \begin{figure}[t]
%     \centering
%     \includegraphics[scale=0.3]{images/top35_newssource_dist.pdf}
%     \caption{Misinformation publishing pseudoscience and propaganda news sources with volume of tweets shared.}
%     \label{fig:misinfo_news_source_categorization}
% \end{figure}
%\end{wrapfigure}

\section{Related Work}
\label{sec:relwork}

Vaccine hesitancy and misinformation on social media and e-commerce platforms has gained much attention in the past few years \cite{cossard2020falling,juneja2021auditing,germani2021anti}. \citeauthor{cossard2020falling} studied the Italian vaccine debate finding echo chambers of anti-vaccine and pro-vaccine groups in 2016 on Twitter, with interaction between the communities being asymmetrical, as vaccine advocates ignore the skeptics. \citeauthor{miyazaki2021strategy} \citeyear{miyazaki2021strategy} recently characterized reply behaviour of anti-vaccine accounts in the COVID-19 vaccine discussion on Twitter, finding that anti-vaccine accounts reply most to neutral accounts using toxic and emotional content. The works have not considered misinformation or conspiracies promoted by anti-vaccine communities, in comparison with our work.

Misinformation and coordinated campaigns during COVID-19 have been highly prevalent throughout the pandemic \cite{sharma2020covid,memon2020characterizing,sharma2021identifying,jamison2020not,lewandowsky2021covid}. Several studies in social science have conducted surveys to identify how vaccine misinformation decreases pro-vaccine intents \cite{enders2020different,jolley2014effects,singh2020first}. \citeauthor{pierri2021impact} \citeyear{pierri2021impact} found  misinformation URLs shared on Twitter are correlated with vaccine hesitancy rates taken from survey data and vaccinations in the U.S., and effect of misinformation on is stronger in U.S. Democratic counties, although hesitancy is higher in Republican counties. \citeauthor{wilson2020social} \citeyear{wilson2020social} found similar correlation with foreign disinformation indicators.

Dataset of English Tweets related to COVID-19 vaccines \cite{deverna2021covaxxy} and longitudinal dataset of accounts promoting anti-vaccine hashtags sampled from Twitter discussions \cite{muric2021covid} have been curated recently by the research community. Lastly, another line of work investigates effects of platform actions on vaccine misinformation. \citeauthor{sharevski2021misinformation} \citeyear{sharevski2021misinformation} examined Twitter's soft moderation efforts against misinformation i.e., warning labels and covers on Tweets with unreliable information, and found that warning covers work but not labels in reducing perceived accuracy of the content, through a randomized survey participant study. \citeauthor{kim2020effects} \citeyear{kim2020effects} looked at YouTube's information interventions on likely misinformation videos, and observed reduced traffic on affected videos.

% \cite{jamison2020not} Vaccine opponents shared the greatest proportion (35.4\%) of unreliable information topics including a mix of conspiracy theories, rumors, and scams. Vaccine proponents shared a much lower proportion of unreliable information topics (11.3\%).

\section{Discussion and Conclusion}

This work examined coordinated campaigns, and other anti-vaccine misinformation and conspiracy communities on Twitter in the context of COVID-19 vaccines discussions. Suspicious coordinated efforts uncovered from the observed account activities, appear to promote a `Great Reset' conspiracy narrative, a Bioweapon conspiracy, with greater automated account-like behaviors. Furthermore, the influence from the anti-vaccine community \emph{and} the far-right conspiracy communities on right-leaning accounts (who retweet top-republican accounts such as Mike Pence) can further distance right-leaning accounts from mainstream and informational health news and science. Misinformation narratives and distortion of COVID-19 vaccine facts is more nuanced as observed through narrative analysis. The structure and differential partisan exposure to COVID-19 anti-vaccine misinformation and conspiracies, coupled with distrust of authorities, and nuanced distortion of true facts especially in the case of COVID-19 vaccines, poses serious challenges to detection/mitigation of this type of misinformation.

We discuss limitations of the work with future research opportunities. First, the study of misinformation is based on unreliable/conspiracy news sources for analysis of misinformation narratives. \citeauthor{Bozarth_Saraf_Budak_2020} \citeyear{Bozarth_Saraf_Budak_2020} evaluated the effect of using low-quality news sources for misinformation in previous studies, and found that the dataset bias affects the prevalence of misinformation in the analysis, but the temporal and narrative differences remain moderately consistent. Second, we estimated correlation of vaccination uptake and misinformation rate, but did not study the causal estimation of misinformation exposure through social media. Few interesting future research questions that arise from the insights in our work, is to study whether social media exposure changes user perception in different partisan communities. In addition, does biased exposure to rarer vaccine side-effects in misinformation narratives cause increased vaccine hesitancy? \citeauthor{jolley2014effects} provides participant studies to show that when participants are shown support for anti-vaccine conspiracies, compared to anti-conspiracy or control groups, they have reduced vaccination intent. The social media based causal studies we proposed here could provide additional evidence and insight for mitigation of anti-vaccine conspiracies.

% \section*{Acknowledgements} 
% This work is supported by NSF Research Grant CCF-1837131. Views and
% conclusions are of the authors and should not be interpreted as representing the social policies of the funding agency, or the U.S. Government.

\fontsize{9pt}{10pt} \selectfont
\bibliography{aaai22.bib}

\end{document}